\documentstyle[10pt,emulateapj]{article}

\def\solm{M$_{\odot}\,$}

\begin{document}     
\vspace{-7cm}
\title{The Luminosity, Stellar Mass, and Number Density Evolution of Field 
Galaxies of Known Morphology from $z = 0.5 - 3$}

\author{Christopher J. Conselice$^{1,2}$, Jeffrey A. Blackburne$^{1,3}$, Casey Papovich$^{4}$}

\altaffiltext{1}{California Institute of Technology, Mail Code 105-24, Pasadena
CA 91125}
\altaffiltext{2}{National Science Foundation Astronomy \& Astrophysics Fellow}
\altaffiltext{3}{Current Address: Physics Department, Massachusetts Institute of Technology}
\altaffiltext{4}{Steward Observatory, University of Arizona}

\begin{abstract}

The evolution of rest-frame B-band luminosities, stellar masses, and number
densities of field galaxies in the Hubble Deep Field North and South are
studied as a function of rest-frame B-band morphological type out to 
redshifts $z \sim 3$
using a sample of 1231 I $< 27$ galaxies with spectroscopic and photometric
redshifts.  We find that the co-moving number, and relative
number, densities of ellipticals and 
spirals declines rapidly at $z > 1$, although  
examples of both types exist at $z > 2$.  The number and number fraction of 
peculiar 
galaxies consistent with undergoing major mergers rises dramatically 
and consistently at redshifts $z > 2$.  Through simulations we argue that 
this change is robust at the 4 $\sigma$ level against morphological 
k-corrections and  redshift effects.  We also trace the evolution of 
rest-frame 
B-band luminosity density as a function of morphology out to $z \sim 3$, 
finding that the luminosity density is steadily dominated by peculiars
at $z > 1.5$ with a peak fraction of 60-90\% at $z \sim 3$.  By $z \sim 0.5$
B-band luminosity fractions are similar to their local values. At $z \sim 1$ 
the B-band luminosity densities of ellipticals and spirals are similar,
with a combined contribution of $\sim$ 90\% of the total luminosity at
$z < 1$. The stellar mass density follows
a similar trend as the luminosity density, with some important exceptions. 
At high redshifts, $z > 2$, 60-80\% of stellar mass appears attached to 
peculiars, while at $ z < 1$, 80\% - 95\% of 
stellar mass is attached to ellipticals and spirals.  The total
integrated stellar mass density of peculiars slightly declines at lower 
redshift, suggesting that these systems evolve into normal galaxies. In 
contrast to the luminosity density,  the stellar mass
density of ellipticals is greater than spirals at $z < 1$, and the stellar
 masses
of both types grow together at $z < 1$, while number densities remain
constant.    From a structural analysis of
these galaxies we conclude that galaxy 
formation at $z > 2$ is dominated by major merging, while at $z < 1$ the 
dominate modes are either minor mergers or quiescent star formation produced 
by gas infall.    Finally, at $z \sim 1.5$
the co-moving luminosity, mass, and number densities of spirals, ellipticals 
and peculiars are nearly equal, suggesting that this is the `equilibrium' point
in galaxy evolution and an important phase transition in the 
universe's history.

\end{abstract}

\section{Introduction}

Understanding how and when nearby galaxies formed and evolved into their 
present physical and morphological states is a major goal of 
contemporary astrophysics.    Most mass in the universe
is likely attached to galaxies, thus understanding the origins of these
systems is central to any ultimate theory of how structure developed.  
Galaxy formation is now at the fore-front of cosmology,
as many major cosmological questions and parameters appear to
be known with some certainty (Spergel et al. 2003).  There is luckily progress
on this front.
The nearby bright galaxy population, defined as galaxies brighter
than M$_{\rm B} < -20$, are becoming fairly well characterized with recent
surveys such as 2dF and the Sloan Digital Sky Survey (SDSS) (e.g., Kauffmann 
et al. 2003; Bell et al. 2003), and will become even more so after the SDSS 
is completed.   Thus far, Sloan and others has argued that $\sim 30$\% of the 
luminosity
in the nearby universe originates from early-type galaxies, with
an even higher fraction of contemporary stellar mass in these evolved systems 
(e.g., Fukugita, Hogan \& Peebles 1998; Blanton et al. 2001; Hogg et al. 
2002; Baldry et al. 2003).  The nearby
bright galaxy population is also clearly bimodal with a large population
of old massive systems, and a large population of less massive
galaxies with younger stellar populations (Kauffmann et al. 2003; 
Baldry et al. 2004).  The morphological distributions of nearby
galaxies is however, not as well studied.
An analysis of the largest morphological catalog of galaxies, the ``Third Reference Catalogue of Bright Galaxies'' (de Vaucouleurs et al. 1991), show that 
approximately 75\% of all galaxies
with M$_{\rm B} < -20$ are spirals, while about 23\% are ellipticals
and the remainder are irregular galaxies (see also Marzke et al. 1994).   
How and when did this
$z \sim 0$ morphological distribution come into place, and how is this
related to the development of stellar populations?  We can answer
these questions directly by looking at galaxies at high redshift with
the Hubble Space Telescope (HST).

Deep ground based and Hubble Space Telescope (HST) imaging
(e.g., Glazebrook et al. 1995; Driver et al. 1995;
Madau et al. 1996; Brinchmann et al. 1998; Steidel et al. 1999; 
Conselice et al. 2003; Dickinson et al. 2003) has
revealed basic evolutionary trends in the evolution of field galaxies.
Lilly et al. (1995) argue from an I-band selected survey that the
luminosity density of galaxies, and thus the star formation rate,
increases as a function of redshift out to $z \sim 1$.  Madau et al.
(1996) followed this up by examining the rest-frame UV luminosity densities
out to $z \sim 4$, demonstrating that the star formation rate remains
high out to these redshifts.  Many of these star forming galaxies
were found to be irregular in appearance to $z = 1$ as shown by HST imaging
of these fields (e.g., Brinchmann et al. 1998).  Complementary to
this work is investigating the stellar mass evolution of galaxies (e.g., 
Dickinson et al. 2003; Rudnick et al. 2003; Fontana et al. 2003; Glazebrook
et al. 2004).   These studies find
that at $z \sim 3$, only approximately 10\% of modern stellar mass is
formed, which rises to $\sim$ 50-75\% at $z \sim 1$.  If these results
are not biased by missing a substantial population of massive galaxies, or
by cosmic variance, then galaxy formation is a gradual process that occurs 
over the age of the universe.   It appears that we are coming to an 
agreement concerning when galaxies in our universe formed and the next step is
determining how this formation occurred.

There are at least three main methods for how galaxies can form: major 
mergers (with mass ratios of 3:1 or lower),
minor mergers (with mass ratios of 3:1 or higher), and gas accretion from the 
intergalactic medium.  Understanding which of these processes is occurring 
requires examining the internal features of galaxies either through high
resolution imaging or integral field spectroscopy.
It has been known for some time that a significant fraction of galaxies in 
the distant universe have irregular/peculiar morphologies which
may be signs of merger activity (e.g., Glazebrook et al. 
1995; Driver et al. 1995; Abraham et al. 1996).   
These early studies were however limited as many did not have redshifts, and 
thus could not trace the cosmic evolution of galaxy structure.  
Recently, van den Bergh et al. (2000), Brichmann \& Ellis (2000),
van den Bergh, Cohen \& Crabee (2001), and Kajisawa \& Yamada (2001) who use 
redshift information, 
argued that there is a real evolution in morphology as a function of
time. Van den Bergh et al. (2000) argued that very few ellipticals and 
spirals exist
at redshifts $z > 1.5$ and other morphological properties seem
to change slowly from $z \sim 1.5$ until today.  
We know that this change in morphology at high redshift is not the 
result of so-called morphological
k-corrections, where galaxies look more irregular at shorter wavelengths
(Windhorst et al. 2002; Papovich et al. 2003).   
Conselice et al. (2003) use HST imaging of the HDF to argue that 
galaxy formation for the most massive systems is dominated by merging and 
calculates the basic merger history of field galaxies.  By uncovering how 
galaxies and stellar mass form as a function of morphological type and 
structure we can begin to determine the formation modes of galaxies.

To perform this analysis we use the Hubble Deep Field North and South
WFPC2 and NICMOS images with the CAS (Concentration, Asymmetry, Clumpiness)
classification system (Conselice
2003) combined with eye-ball morphology, color, stellar mass, and 
luminosity information.  In this paper we characterize the 
evolution of stellar masses, luminosity densities, and number densities as a 
function of morphology.  We show that
at the highest redshifts most stellar mass is attached to peculiar
galaxies.  At lower redshifts the amount of stellar
mass in peculiars gradually decreases while the amount in ellipticals and
spirals gradually increases, verifying on a longer time-scale, the results
of Brinchmann \& Ellis (2000) who find a similar trend for
galaxies at $z < 1$. One basic result is that at $z \sim 1.5$ the
co-moving densities, co-moving B-band luminosities and co-moving stellar
mass densities are nearly equal for peculiars, disks, and ellipticals.  This
is the same epoch where disks and ellipticals first appear morphologically
in abundance.  At redshifts $z > 1.5$ peculiar galaxies, which
we argue based on the CAS systems are mergers, increasingly dominate  
number, luminosity, and mass densities out to $z \sim 3$.  At $z < 1.5$, we 
find, similar
to Brinchmann et al. (1998) and Brinchmann \& Ellis (2000), that ellipticals
and disks dominate in number, luminosity and stellar mass density. 
 We further conclude that major mergers
are the dominate mode for forming the most massive galaxies at $z > 1.5$  
while minor mergers and gas accretion from the intergalactic medium are the
likely cause of star formation and stellar mass assembly at $z < 1.5$.

This paper is organized as follows:
\S 2 is a description of the data used in this paper and
the morphological techniques used to study galaxies in the
two Hubble Deep Fields.  \S 3 is the description of our analysis while
\S 4 \& \S 5 are a discussion and a summary of our results.  We assume 
a cosmology with H$_{0}$ = 70 km s$^{-1}$ Mpc$^{-1}$, 
$\Omega_{\Lambda}$ = 0.7 and $\Omega_{\rm m}$ = 0.3 throughout.

\section{Data and Analysis Method}

\subsection{Images}

Our basic imaging data are the Hubble
Deep Field North WFPC2 (Williams et al. 1996) and NICMOS fields
(Dickinson et al. 2000, 2003) (hereafter HDF-N) as well as the Hubble
Deep Field South WFPC2 images (hereafter HDF-S) (Casertano et al. 2000) and
the VLT near infrared ISAAC images of the HDF-S (Labb{\'e} et al. 2003).   
In these two fields, we use
the following broad-band filters: F450W (hereafter B$_{450}$), F606W (hereafter
V$_{606}$), F814W (hereafter I$_{814}$) from WFPC2, and F110W 
(hereafter J$_{110}$) and F160W 
(hereafter H$_{160}$) from NICMOS for the HDF-N.  The ISAAC data for
the HDF-S includes J$_{\rm s}$, H and K$_{\rm s}$ band imaging over a 2.5\arcmin $\times$ 2.5\arcmin\,
area to 3$\sigma$ AB magnitude depths of 26.8, 26.2 and 26.2, respectively.
These data are used 
for the structure analyses, and for deriving 
photometric redshifts and other photometric and structural properties
that we use throughout this paper.  Ground based infrared photometry for both 
the HDF-N and HDF-S galaxies are also included in these measurements (e.g,.
Dickinson et al. 2003).

\subsubsection{Catalogs of Galaxies}

The galaxies in the two Hubble Deep Fields were selected using SExtractor
(Bertin \& Arnouts 1996).  For the HDF-S the detection was done on the 
I$_{814}$ band image, while a combined J$_{110}$+H$_{160}$ image was used for
the HDF-N.  We then match these catalogs to all further data, including
photometric redshifts.  We also use these catalogs for our morphological
analyses.
From these catalogs, a total of 1288 objects were detected in the Hubble Deep
Field North and 1293 in the HDF-S.  We classify all of these 2581 objects,
in the two HDFs, although only a fraction have meaningful morphological
parameter measurements.    To avoid problems associated with fainter and
less resolved galaxies we do a cut on galaxies to remove
systems with apparent magnitudes
I$ > 27$ from our analyses.  Galaxies fainter than this 
generally cannot have their
morphological parameters measured with any certainly, due to the
their low-S/N and resolution (Conselice et al. 2000a, Conselice 2003; 
\S 2.3).  

\subsubsection{Redshifts - Spectroscopic and Photometric}

There are spectroscopic 
redshifts available for galaxies in the HDF-N, with
the bulk from the surveys of Cohen et al. (2000) and Steidel et al. (2003), but
other sources as well.  These redshifts are typically for galaxies brighter
than $R = 25.5$, although some fainter sources have redshifts. These
redshifts are discussed in detail in Budav\'ari et al. (2000).
For the HDF-S, there are 43 redshifts from the VLT study of Sawicki \& 
Mallen-Ornelas (2003).  The Sawicki \& Mallen-Ornelas galaxies are all at 
$z < 1$ however. For the remaining galaxies in 
these fields, we use photometric redshifts as derived by
Budav\'ari et al. (2000) for the HDF-N galaxies, and by the Stony Brook
photometric redshift group (e.g.,  Yahata et al. 2000) and
the FIRES team (e.g., Rudnick et al. 2003) for the HDF-S fields.  The
photometric redshifts for Rudnick et al. (2003) are however limited in
number and we use the Stony Brook redshifts whenever there is no
spectroscopic redshift or FIRES photometric redshift.
The accuracy of the photometric redshifts for the HDF-N are discussed in
detail in Budav\'ari et al. (2000). For the HDF-S, no comparison between
the photometric redshifts of the Lanzetta group and the spectroscopic
redshifts have yet been done.  Figure~1 shows this comparison for
galaxies with spectroscopic redshifts in the HDF-S field.  
The scatter between the photometric redshift and the spectroscopic
redshifts is $\sigma$ = 0.08 with an average offset of 0.06 and an
average ($\delta z/z) = 0.11$.  This is a fairly good accuracy, and suggests
that the other photo-zs are likely reliable, on average, within this 
uncertainty. However as photo-zs are less accurate at fainter magnitudes,
they are likely less reliable for less luminous and/or more distant galaxies. 
We compute rest-frame properties of these galaxies, including
their absolute blue magnitudes (M$_{\rm B}$, rest-frame colors, and sizes) 
using these redshifts (see also Dickinson et al. 2003).

\subsection{Photometric Properties}

Throughout this paper we use the broad-band HST and ground based
photometry to determine rest-frame absolute magnitudes, colors,
and sizes for our sample.  The photometric properties of
galaxies are determined through interpolating rest-frame values
of M$_{\rm B}$ and $(B-V)$.  No K-corrections are computed as
we directly measure these values from the SED and knowledge of
the redshift for each galaxy.  See Dickinson et al. (2003) and
Conselice et al. (2003) for more details on how these interpolations
are done.

\subsection{Morphological Classifications}

We performed eye-ball classifications for all 2581 galaxies in the observed
I$_{814}$ band, as well as in the H$_{160}$ for the HDF-N.  Based on
these images a classification was made.  Each galaxy
was put into one of six categories: star, elliptical, early-type
disk, late-type disk, edge-on disk, irregular, merger/peculiar, 
and unknown/too-faint.
These classifications are very simple and are only
based on appearance.  No other information, such as colors or redshifts
were used when determining types.  We discuss in \S 3.1 biases
due to redshift effects and morphological k-corrections. Ellipticals, 
early-type disks and
late-type disks were classified based on the presence of apparent bulge and
disk components through the following criteria:

\begin{enumerate}

\item {\bf Ellipticals:} If a galaxy appeared to be centrally concentrated 
with
no evidence for a flatter, lower surface brightness, outer structure it was 
classified as an elliptical.  

\item {\bf Early-type disks:} If a galaxy contained a central concentration 
with some
evidence for lower surface brightness outer light, it was classified as
an early-type disk. 

\item {\bf Late-type disks:} Late-type disks are galaxies that appear to have
more outer low surface brightness light than inner concentrated light.

\item {\bf Irregulars:} Irregulars are galaxies that appear to have no 
central light concentration and a diffuse structure, sometimes with clumpy 
material present.  

\item {\bf Peculiars:} Peculiars are systems that appear to be disturbed
or peculiar looking.  These galaxies are possibly in some phase of
a merger (Conselice et al. 2003) and are common at high redshifts.

\item If a galaxy was too faint for any reliable classification it was placed
in the {\bf unknown/too-faint} category.   Often these galaxies appear 
as smudges without any structure.   These could be disks or
ellipticals, but their extreme faintness precludes a reliable classification.
Often, a system appeared very faint, but large scale peculiarities, such
as warping and non symmetrical structure could be detected by eye. 
If any of these systems displayed
this peculiarity it was placed into the peculiar category.

\end{enumerate}

The usefulness of this type of morphological classification is questionable
(e.g., Conselice 2003) since Hubble types, and other eye-based classifications,
are subjective.  There is also the issue that fainter galaxies
will be more difficult to classify, although classifications in the HDFs
are on average possible to I = 27 (Conselice et al. 2000a).   This is fainter
than the limit used in Abraham et al. (1996), whose use a I = 25 limit.
Our analyses are based on definitive 
classifications of broad types, with a generous 'unclassifiable' option, and 
these should not be 
affected significantly by morphological classification biases.   It is clear
from the data that to I = 27 classifications can be performed for
the majority of galaxies.  Between I = 26 and I = 27, 38\% of systems
are unclassifiable, this rises to 61\% between I = 27 and I = 28.  We thus
use the limit I = 27 for obtaining a morphological classification for
the majority of galaxies in the HDFs.   To test the reliability of
this classification, we do a series of simulations of nearby
galaxies (\S 2.6.1) and HDF-N galaxies between $z \sim 0.5 - 0.8$
placing them at higher redshifts.  When we reclassify these galaxies after
simulation we find that most ($\sim$ 70\%) with magnitudes $I = 26 - 27$ 
are still classifiable by eye into their original types, justifying
our use of this magnitude limit.  Note that at $I = 27$, quantitative 
classifications
are also possible, but require corrections due to surface brightness dimming
(Conselice et al. 2000a; \S 2.6.1).  We also repeat
this eye-ball classification for the HDF-N using the NIR H$_{160}$ band images
to test for effects of morphological k-corrections (\S 2.6) and to obtain a
rest-frame B-band classification for the HDF-N sample.

\subsection{CAS Parameters}

To quantify the structural features of galaxies in the
HDFs, we use the concentration, asymmetry and clumpiness (CAS)
system described in detail in Conselice (2003).  In the 
CAS system the concentration index ($C$) is a measure of how centrally
concentrated or compact a galaxy is (Bershady et al. 2000).  Based on
empirical correlations the concentration index
is a measure of the form, and scale of a galaxy (i.e., its 
mass).  Concentrations correlate
fairly well with the absolute magnitude of a galaxy and its bulge
to disk ratio (Conselice 2003).    
The asymmetry index ($A$), first measured by Schade et al. (1995), 
Abraham et al. (1996) and Conselice (1997) is a measure of the global 
distortion in a galaxy 
from a completely symmetric structure, and is a good method for indicating
galaxies undergoing major mergers (Conselice et al.  2000a, 2000b;
Conselice et al. 2003).  The clumpiness
index, $S$, correlates with the amount of star formation in 
a galaxy, as shown by its fairly strong correlation with H$\alpha$ fluxes
and broad-band colors (Conselice 2003).   There are many issues concerning
how these parameters are measured, which are discussed at length in
Conselice et al. (2000a), Bershady et al. (2000) and Conselice (2003).

\subsubsection{Contamination from Galaxies - Segmentation Maps}

Since the CAS system is model independent and works in batch mode, the
catalogs for the various HDFs were fed into an IRAF based program that
computes CAS parameters and Petrosian radii for
each galaxy (see Conselice 2003).  Before we ran any of the CAS programs, we 
performed
several pre-analysis quality control steps.  The most important of
these was to minimize as much as possible contamination from neighboring
galaxies. This was done was by creating SExtractor (Bertin \& Arnouts
1996) segmentation maps that reveal unambiguously pixels belonging
to each galaxy.  These maps are created during the SExtractor detection
process.   The form
of the segmentation map is such that every pixel is valued either zero or
has a integer value, corresponding to a catalog entry, that reveals which
galaxy that pixel belongs.  Based on these maps, we isolated
each galaxy's segmentation area and placed all light from other galaxies
to the background level and noise before computing the CAS parameters.
Systems which are merging may be artificially separated by this process
into two separate galaxies. This is largely not a problem however as
galaxies which are undergoing mergers are asymmetric with, or without,
the presence of the other galaxy.  The only concern about these
galaxies is that they might be double counted when considering the number of
mergers (see Conselice et al. 2003).  We are also not biasing asymmetry
measurements by replacing the light from nearby galaxies with the background,
and therefore potentially not retrieving the underlying light from the galaxy
being measured. The reason for this is that the asymmetry computation is
heavily weighted towards the brightest parts of a galaxy, and any faint outer
light contributes very little to the measurement (Conselice 2003).

There are many other technical issues that are involved in applying an 
automated and quantitative approach such as measuring CAS parameters.  
These issues include correcting for noise and defining the radius and
center by which these parameters are measured.  The noise in the HDF-N
and HDF-S WFPC2/NICMOS images is enhanced near the edges of
the frames, and thus we remove these systems from our analyses.  
To deal with noise in a general sense we follow the
procedure outlined in Conselice et al. (2000a, 2003) for understanding and 
accounting for systematics presented by the background.  We
further follow the techniques in these papers when accounting for radii and 
centering routines when computing the CAS parameters.  We give
only a brief description of these procedures here and refer the reader
to previous papers for
detailed explanations for how CAS parameters are computed.

\subsubsection{Background Removal}

One critical component to understanding and correct for when
measuring CAS parameters 
is the presence of background noise.  This noise originates from read noise,
photon counts from background light, and any intrinsic variations of
this background within a small sky area.   Our background correction
method as outlined in Conselice et al. (2003), will correct for both
correlated and uncorrelated noise, assuming that in either case it is also
present where the galaxy itself is found.  As noted
in Conselice et al. (2003) there are several problems to this
methodology, most importantly the fact that the sky varies across these
images, particularly in the NICMOS bands (Conselice et al. 
2003).  We obtain a partial solution to this problem by computing the CAS
parameters using several background areas across each WFPC2 and NICMOS image.
This gives us an understanding of how
the background light contributes to errors in our measurements, as
each CAS run gives us a slightly different set of parameters
depending on the background area chosen.  After measuring CAS
values using these different backgrounds we find
that the statistical variation in the computation of these parameters
across images is smaller than the measurement errors.  This effect therefore
does not contribute significantly to the error budget of our CAS 
measurements.

\subsubsection{Radii}

We use in this paper the 1.5 $\times$ r$(\eta = 0.2)$ inverted
Petrosian radius described
and calibrated in detail in Bershady et al. (2000), Conselice et al.
(2000a) and Conselice (2003) to measure CAS parameters.  Briefly, the 
inverted Petrosian radius
is the radius where the surface brightness in an annulus at that radius
is a given fraction of the surface brightness within the radius.  Our
use of $\eta = 0.2$ to define the radius of our galaxies is the same
as that used by the Sloan Digital Sky Survey to measure magnitudes 
(e.g., Blanton et al. 2001).
Based on simulations, it can be shown that for most popular galaxy light
profiles, such as exponential, gaussian, and de Vaucouleur, the Petrosian
radius, defined this way, contains $\sim 99$\% of the light from these systems.
The CAS program also measures the half-light radii of these galaxies, within
the 1.5 $\times$ r$(\eta = 0.2)$ radius.

\subsubsection{Rest-Frame B-band CAS Values}

Using the wavelength coverage of the WFPC2 bands, we
are able to determine the rest-frame B-band morphologies of our sample 
in both the HDF-N and HDF-S out to $z \sim 1.2$.  We can further compute
rest-frame B-band CAS parameters out to $z \sim 2.5$ using the NICMOS
HDF-N imaging.  We compute rest-frame B-band CAS parameters and radii
by linearly interpolating between the values measured in the observed
HST wavebands,
that straddle the observed wavelength of the B-band, at a given redshift.
For galaxies at $z > 2.5$, in the HDF-N, we use the observed H$_{160}$ band 
morphologies and radii.

\subsection{Stellar Masses}

The stellar masses we use are from the analyses of Papovich et al. (2001), 
Papovich (2002)
and Dickinson et al. (2003) for the HDF-N.  We calculate the stellar
masses for the HDF-S using photometry from WFPC2 optical imaging and
infrared imaging form the FIRES survey (e.g.,
Labb{\'e} et al. 2003)  using the techniques
from Papovich et al. (2001).  The stellar masses in the north field
are computed through
spectral energy distributions using Hubble Space Telescope WFPC2 imaging
 in the U$_{300}$, B$_{450}$, V$_{606}$ and I$_{814}$ bands,
as well as NICMOS imaging in the J$_{110}$ and H$_{160}$ bands.  A deep
K-band image of the HDF-N taken with the Kitt Peak National Observatory 4 
meter is
also included.  To determine the stellar masses of the galaxies in both
the north and the south their SEDs were fit by various star formation 
histories with
different metallicities and internal extinction values derived from
Bruzual \& Charlot (2003) stellar population synthesis models.  
After a best fit to the observed SED is found, the M/L ratio is derived and the
stellar mass is computed based on this.  The major parameters for
the fitting are the age of the stellar population and the star formation
time-scale.   There are several different possible
initial mass functions and star formation histories that can
be used to do this modeling of M/L ratios. We however use models where
the IMF is Salpeter and ranges from 0.1 to 100 \solm, the metallicity is 
solar, and the star formation history monotonic. We also use similar
SEDs used to computed the k-corrections for the photo-zs discussed 
in \S 2.1.2.   The best model is then selected by the $\chi^{2}$ technique
with Monte Carlo simulations employed to determine the 68\% confidence or
error intervals on the stellar masses after randomly including changes
within the photometry error budget.  For more details on this process
see Papovich et al. (2001) and Dickinson et al. (2003).

\subsection{Biases}

\subsubsection{Redshift Effects and Morphological K-Corrections}

It is important to understand what effects redshift will have on the
quantitative values of our galaxy classifications, including the
CAS parameters, particularly if redshift
creates a systematic effect of either lowering or rising the threshold for
placing an object into one of the Hubble sequence bins.  This effect has 
already been explored by
Conselice et al. (2000a), Conselice (2003) and Conselice et al. (2003)
although we re-examine this in more detail here in a empirical way here
and in \S 3.1.1.  There are two major effects produced by redshift - the
dimming of galaxies and the morphological k-correction.

To determine how various normal galaxy morphologies change solely due to 
redshift
effects, that is surface brightness dimming, we simulate rest-frame optical 
images of nearby Hubble types from
the Frei et al. (1996) digital catalog of galaxies (see also Conselice
2003) by placing them at higher redshifts. By doing this we are able to
determine how the morphological properties of galaxies change solely due to
redshift.  
We analyze these simulations by performing the exact same CAS runs on these
simulated galaxies as we did on the galaxies in the HDFs themselves.  Thus,
by comparing the CAS parameters for the $z \sim 0$ sample to how their
values change at $z > 1$, we can determine the effects of redshift on
the measurement of CAS parameters, which we can compare to the observed 
evolution to 
determine the true change in CAS values from low to high redshift.
We find that the most significant changes occur at redshifts $z > 1.5$,
particularly for the most asymmetric and clumpy galaxies.  Table~1
lits the changes in the CAS parameters at different redshifts.

Galaxies also look different at different wavelengths, which is a result of 
the so-called morphological k-correction (Windhorst et al. 2002; Papovich
et al. 2003).  Generally, galaxies appear as later types when viewed at 
bluer wavelengths, as star formation is more visible and older
stellar populations are fainter at shorter wavelengths.  Although
we generally use rest-frame B-band morphological classifications in this
paper, there could be some biases, particularly at high redshifts
where the rest-frame I$_{814}$ band samples the rest-frame UV for these
systems.  Figure~2 plots classifications in the H$_{160}$ and I$_{814}$
bands versus each other. The line through the middle overlaps those systems 
which have the same classification in both bands. 

Figure~2 demonstrates that there are some morphological biases between
the two bands.  The most common disagreement is that there are more systems
classified as 'too faint' in the H-band than in the I-band.  This 
results from these systems, which are most commonly ellipticals and
peculiars, being too faint in the H$_{160}$ band due to the shorter
exposure time used to obtain the NICMOS imaging (Dickinson et al. 2000).  
Otherwise, the most common misclassifications are what we would expect. 
For example, there are 13 systems classified
as an early type spiral (eS) in the I$_{814}$ band which are classified
as elliptical in the H$_{160}$, and there are nine systems classified
as a late type spiral (lS) in I$_{814}$ which are classified as early-type
spirals in H$_{160}$.  Most of these systems are at low redshift, where
disks and ellipticals are found in abundance.  We are thus not likely
biased by using the I$_{814}$ band in determining classifications, particularly
peculiars, which remain peculiar at all wavelengths.

\subsubsection{Galaxy Selection}

There are many selection effects which we discuss throughout this paper. 
The most obvious is that we use a I$_{814}$ $< 27$ magnitude selection for 
galaxies in our analysis.  This
selection allows us to determine with accuracy the morphologies of
galaxies under study, but removes optically faint
galaxies at high redshift from our sample.  As such, it is important to 
determine
the characteristics of this selection for our analysis. The first step in
understanding biases from the I$_{814}$ $< 27$ selection is to determine 
the redshift distribution of our I$_{814} < 27$ galaxy 
sample in both the HDF-N and HDF-S (Figure~3). As can be
seen, the peak in the distribution is at $z \sim 1$, with a smaller peak at
higher redshifts around $z \sim 2$. 

To determine how faint we can go in our analysis without
missing objects, we investigate how absolute magnitudes vary with
redshift (Figure~4).  At redshifts $z < 2$ all galaxies brighter
than M$_{\rm B} = -20$ appear to be detected without a significant 
incompleteness.  As we sometimes use this magnitude limit in this paper, we
can draw global morphological conclusions at redshifts $z < 2$ for galaxies
with $M_{\rm B} < -20$.  At $z > 2$, the I$_{814} < 27$ limit is likely 
incomplete and we cannot be confident that we are not missing some fraction
of the galaxy population.
Unless otherwise stated, we use the I$_{814} < 27$ limit in our analysis.

The fact that we are not missing galaxies at $z < 2$ can also
be seen in another way by examining Figure~5 which plots the
apparent magnitudes of all HDF-S and HDF-N galaxies, as a function of 
redshift, out to $z \sim 3$ with the I$_{814} < 27$ limit imposed.  
The solid, short dashed and long dashed lines are the expected 
apparent magnitude of M$_{\rm B} = -20$ ellipticals, Sa and Sc galaxies,
respectively, as 
a function of redshift.  The value M$_{\rm B} = -20$ is roughly the
value of L$^{*}$ for modern galaxies. 
The apparent magnitudes were computed through
the formula: I$_{814}$ = M$_{\rm I}$ + DM + K, where M$_{\rm I}$ is the
absolute I-band magnitude of each type computed from Fukugita et al. (1995),
DM is the distance modulus, and K is the K-correction.  K-corrections for
each type were computed from SEDs presented in Poggianti (1997).  No
evolutionary corrections have been applied to these magnitudes, which would 
only make the galaxies brighter, especially the ellipticals.  As a note,
the recently discovered evolved massive galaxies in the HDF-S by Franx
et al. (2003) all would be included in our analysis as they have
magnitudes $I < 27$, and are very bright in optical magnitudes with
M$_{\rm V} = -23$ to M$_{\rm V} = -25$ (van Dokkum et al. 2003) and
with M$_{\rm B} < -20$.

\section{Analysis}

\subsection{The Evolution of Number Densities}

The first cut at a basic morphological analysis involves examining how
our eye-ball estimates of galaxy types are distributed with redshift.
Examining the raw numbers of ellipticals and spirals as a function of type 
reveals a deficit of ellipticals and spirals
at redshifts larger than $z \sim 1.5$ (Figures 6 \& 7).  There is
a gradual decline in the numbers and relative numbers of
normal galaxies at $z > 1$ as peculiar galaxies become the dominate
population.
This change remains when we redo the entire classification 
independently using the H$_{160}$-band images of the HDF-N, which allows 
us to sample the rest-frame optical to $z \sim 2.5$.  Thus, this change is not 
solely due to observing galaxies in the rest-frame ultraviolet at $z > 1.5$ 
where galaxies can looked different than in the optical 
(Windhorst et al. 2002).  
Previous studies (e.g., van den Bergh et al. 2001) were not able to make 
this claim.
 
To argue that this change occurs, which we call the galaxy structure-redshift
relationship (Conselice 2004), we compute the number of ellipticals and 
spirals in equal co-moving volumes of $\sim 1131$ Mpc$^{3}$ arcmin$^{-2}$ 
(Figure~6) for both a I$ < 27$ limit and a M$_{\rm B} < -20$ limit.
Figure~6 shows that there appears to be a gradual drop in the co-moving density
of both spirals and ellipticals at higher redshifts for both
selections.  The computed co-moving densities for the I$< 27$ sample
are listed in Table~2. Note that this drop is
seen in both the HDF-S and HDF-N, thus it is unlikely to be purely the
result of cosmic variance.  The drop-off in number becomes more pronounced
when we consider only those systems that are brighter than M$_{\rm B}
= -20$. Note that this change is gradual, and does not occur instantaneously,
but is most rapid between $1 < z < 2$.

This can be seen further from Figure~7 which shows the relative fraction of 
ellipticals, spirals, peculiars,  and galaxies too faint for classification, 
and how their relative densities evolve as a function of 
redshift.  This figures remains largely identical when using the
M$_{\rm B} = -20$ limit.   The most notable trend
in Figure~7 is the gradual increase
in the fraction of galaxies with peculiar morphologies at higher redshifts,
with a corresponding decrease in the fraction of ellipticals and
spirals.     There is also a general relative increase with redshift
in the fraction of galaxies too faint for classification in all three images.
From Figures~6 \& 7 it also appears that there
are galaxies that morphologically appear as disks and ellipticals
at redshifts $z > 1.5$, although the numbers are small (cf. Conselice
et al. 2004).  Finally, we note that at $z \sim 1.5$ the relative
number density of ellipticals, disks and peculiars is relatively similar.

\subsubsection{Is the decline in Hubble Types at $z > 1.5$ real?}

The decline in Hubble types at $z > 1.5$ has been
noted before (van den Bergh et al. 2001; Kajisawa \& Yamada 2001), but its 
suddenness deserves further attention and a careful analysis. One aspect of 
this that has not been examined is how much of the drop in normal galaxies
is due to redshift effects.
To answer this question we simulate galaxies that are
at z$\sim 0.65$ in the HDF to how they would appear at $z \sim 1.5$.
This simulation is done following the procedure outlined in Conselice
(2003) and Conselice et al. (2003). In summary, the procedure is to take the 
observed V$_{606}$-band HDF-N image, which corresponds to
rest-frame B-band for objects at $z \sim 0.65$, and reduce it in resolution 
and signal to noise to match how these $z \sim 0.65$ galaxies would appear
in NICMOS images of the Hubble Deep Field North if they were at 
$z \sim 1.5$.  By doing this experiment we can determine what fraction of 
spirals and ellipticals are identifiable at this redshift.  Note that we 
use the NICMOS HDF-N H$_{160}$ observational
details for this simulation as this filter samples rest-frame optical light
at $ z \sim 1.5$.
After carrying out this simulation, we re-classified in the same manner as 
described in \S 2.3 all galaxies originally
in the redshift range 0.5 - 0.8, simulated to how they
would appear at $z \sim 1.5$.  Between redshifts $z \sim 0.5$ and $0.8$ 
in the HDF-N there are 130
galaxies brighter than M$_{\rm B} = -19$, of which 35 are classified as 
ellipticals and 44 as spirals.  When we
perform the $z \sim 1.5$ simulation and reclassify all objects, we
find that 28 of the ellipticals and 29 of the spirals are still identifiable
as such.  This gives us a detection fraction, f$_{\rm det}$, of 0.8 for
ellipticals and 0.66 for spirals.
  
To determine the significance of the drop in ellipticals and
spirals at $z \sim 1.5$ we first must determine the systematic effects
of redshift on the detectability of these simulated galaxies.  Since the
drop off in Hubble types occurs at roughly $z \sim 1$, we take the
average number, per co-moving Gpc$^3$, of ellipticals and spirals at
redshifts $z < 1$.  We then multiply these densities by the detection
fraction from our simulation to determine the number of spirals and
ellipticals we would expect to find at $z \sim 1.5$, if no morphological,
luminosity, or number density evolution were occurring.  The numbers 
in our fiducial $\sim$ 1131 Mpc$^{3}$ co-moving volume we expect
to find at $z \sim 1.5$ are $\sim 24$ ellipticals and 
$\sim 26$ spirals.  Since we find only three ellipticals and
three spirals in this redshift range, there is clearly a significant
difference between the number of Hubble types expected and those
found.  The formal significance of this drop is 4.2 $\sigma$ for
the ellipticals and $\sim 4.6$ $\sigma$ for spirals.  As we have
considered all redshift effects: morphological k-corrections, dimming
and resolution, in this calculation we can conclude with confidence that
the drop in Hubble types seen at $z \sim 1.5$ is likely real.

The gradual increase of peculiar galaxies at the expense of normal Hubble
types is circumstantial evidence that at least some fraction of modern
spirals and ellipticals originate from peculiar galaxies (see Conselice
et al. 2003; cf Brinchmann \& Ellis (2000) for objects at $z < 1$).   
There appears to be a relatively similar number of spirals
and ellipticals at $z < 1$ in both the HDF-N and HDF-S.
At redshifts $z \sim 1.5 - 2.5$ where the H$_{160}$ data is sampling
the rest-frame optical light from these galaxies, we sill find that the
fraction of peculiars increases at the expense of Hubble types.
The galaxy population is however evenly distributed in co-moving
number densities for all types at $z \sim 1.5$.  This appears
to be the equilibrium point between the number of spirals, ellipticals
and peculiars, which breaks down at higher and lower redshifts.

We see similar number density trends in the Hubble Deep Field North and 
South, although
we must deal with the issue of cosmic variance.
Having two field effectively reduces the effects of cosmic
variance by a factor of two (Somerville et al. 2004).  It is difficult to
quantify this for the different galaxy types in the HDFs which have different
clustering scale lengths.  We can investigate this for the Lyman-Break galaxies
in the HDFs however.  The relative cosmic variance (Somerville et al. 2004) 
for LBGs in one
HDF has a $\sigma_{\mu}^{2}$ = 40\% and 20\% when using the two fields.
The drop in Hubble types at $z > 1$ has
also been seen in other deep HST fields, such as the CFRS fields (van
den Bergh 2001).  Future studies, such as the GOODS survey (Giavalisco
et al. 2004) will better address the cosmic variance problem.

\subsection{Luminosity Evolution}

\subsubsection{Luminosity Functions of Different Morphological Types}

The luminosity functions of galaxies can reveal important evolutionary
trends (Ellis 1997) as has been measured fairly well out to $z \sim 1$ 
(e.g., Brinchmann et al. 1998; 
Wolf et al. 2003).  The luminosity function has been further
measured out to $z \sim 1$ as a function of color (e.g., Lilly et al. 1995),
and morphological type (Brinchmann et al. 1998).  
How the luminosity function evolves with morphology, and
the resulting evolution in the co-moving density as a function of morphology,
reveal in which types of galaxies star formation is occurring in, and when
transitions to other morphologies occur.  In Figures~8 \& 9 we present the
total luminosity function for galaxies in the HDF-North and South separated
into redshift bins and morphological type (ellipticals, disks, peculiars, 
and galaxies too faint for reliable classifications.)  We parameterize the 
luminosity function using the 1/V$_{\rm max}$ formalism (Schmidt 1968; 
Felten 1976):

$$\phi({\rm M, morph, z}) {\rm dM} = \Sigma \frac{1}{\rm V_{max}}$$

\noindent where the summation is over each galaxy in a given magnitude 
(M), morphological-type, and redshift (z) bin.  The result of these luminosity
functions are shown for the Hubble Deep Field North and South in
Figure~8 by crosses (HDF-S) and circles (HDF-N).  The value
of V$_{\rm max}$ is the maximum co-moving volume such that a particular
galaxy would be observed within the I $< 27$ sample definition and using
k-corrections from Poggianti (1997) determined by the morphological type of
each galaxy.  The range of these k-corrections in the I-band can be seen
in Figure~5.  The uncertainties in these luminosity functions are computed by 
Poisson statistics of the number counts, and coupling this
number density uncertainty with the average V$_{\rm max}$ within
a given magnitude, redshift, and morphology bin.

\subsubsection{Luminosity Evolution as a Function of Morphology}

The luminosity density evolution of the universe is used to determine the
evolution in the star formation rate with redshift (e.g., Lilly et al.
1995; Madau et al. 1998).   Using the information in Figures 8 \& 9
we can determine the total luminosity function (LF) and decompose it
as a function of morphology and redshift.  Note that we do not do luminosity
functions corrections as we are interested
in matching luminosities with our classifications.  Figures 8 \& 9 show
that the total B-band luminosity function does not appear to change
drastically at different redshifts except at the bright end. The faint end of
these luminosity functions also tend to match
the 2dF B-band luminosity function (Norberg et al. 2002).  Also, the HDF-N
and HDF-S LFs are nearly identical within their uncertainties, thus we
are unlikely biased strongly by cosmic variance effects or photometric 
redshift errors.

What are the types of galaxies in which B-band luminosity originates? 
This is a fundamental question since
it tells us the morphological and evolutionary state of galaxies
when their stars are created.  Brinchmann et al. (1998) used HST
imaging of CFRS, LDSS and Groth strip fields to argue that there is
a gradual decline in the luminosity density attached to peculiars
since $z \sim 1$ and a corresponding rise in the luminosity density
attached to disks and ellipticals.  Using our smaller, but deeper, HDF
pointings we can take this a step further and perform a similar analysis
out to redshifts $z \sim 3$.  The luminosity function $\phi$ as a function 
of morphological type is shown in Figure~9 while the total luminosity 
densities, and the fraction of total luminosity, as a function of type and
redshift, are shown in Figure~10.  The points
at $ z < 0.9$ on Figure~10 are all taken directly from Brinchmann et al. 
(1998). These data are listed in Table~3 as a function of redshift.
The luminosity densities shown in Figure~10 are computed by adding up the
rest-frame B-band light from the galaxies detected for each morphological
type.  We do not integrate fitted luminosity functions to compute the 
integrate light, or extrapolate to fainter magnitudes based on a fitted
luminosity function.  Doing this for the different morphological
types would not effect the results in any substantial way.

Based on Figures 9 \& 10 it appears that the luminosity density is dominated
at $z > 2$ by galaxies in a peculiar phase.
At $z < 1$ the luminosity density is dominated by normal galaxies, that
is, spirals and ellipticals.  There are a few other effects that can
be determined from Figure~10.  First, the B-band luminosity density of 
peculiars drops from $z \sim 3$ to 0 and rapidly becomes a minor
contributor to the luminosity density at lower redshifts in both the
north and the south HDF.  Furthermore, the 
luminosity density of peculiars and normal galaxies is nearly equal
at $z \sim 1.5$ when normal galaxies begin to appear in the same
number densities as at $z \sim 0$.  

The peak in the
B-band luminosity density at $z \sim 1$ is due to spirals and ellipticals -
not the peculiars.  In fact, at $z < 1.5$ ellipticals
and spirals dominate the B-band luminosity density of the universe.  
Since B-band luminosity traces
star formation as well as existing stellar mass, understanding
the implications for this is not straightforward, but can be
understood better by examining the growth of stellar mass in
these systems (\S 3.3).  This B-band luminosity is however originating
at $z < 1.5$ from a combination of evolved stellar populations plus
new star formation. 
This can further be seen by studying the U-band luminosity density of
objects as a function of redshift. Based on this, it appears that all
galaxy types are contributing to the star formation density (e.g.,
de Mello et al. 2004), that is, not only galaxies undergoing bursts of
star formation, but also normal galaxies, are bright in the U and B bands.

 There
is a second peak at higher redshift $z \sim 2.5$, particularly in the HDF-S, 
which
is dominated by emission from peculiar galaxies.  In the HDF-S there
are several very bright galaxies at $z > 2$ that we identify by eye as 
ellipticals and peculiars. These systems also have large
stellar masses (\S 3.3), and their existence  is one of
the fundamental differences between the two HDFs. It must also be
said that their existence, for the most part, relies on photometric redshifts
that can differ between different groups.  Spectroscopic followup is
necessary to confirm such a large different between the two HDF fields.

\subsection{Stellar Mass Evolution}

The build up of a galaxy's stellar mass occurs when
baryonic gas is converted into stars after cooling. 
When and how this star formation occurs is a fundamental question in
cosmology with most approaches either tracing the star formation directly
(e.g., Madau et al. 1998) or through integrated effects traced
by the stellar masses of galaxies (Dickinson et al. 2003).  
It appears
that over half of all stellar mass in the universe was formed
between $z \sim 1$ and $z \sim 3$, a time interval of $\sim 2.5$ Gyrs.
The modes of star formation during this epoch are important for understanding
how galaxies formed.  Using morphological information, including the CAS
parameters for these galaxies (\S 3.5), we can possibly determine the modes 
by which star formation is occurring.

To determine in which galaxies stars  formed 
we examine the distribution of stellar mass as a function of
morphological type at redshifts from $z \sim 0.5$ to $z \sim 3$.  
Figure~11 shows the total mass functions of galaxies at our four redshift
ranges.  We compute these mass functions based on the same V$_{\rm max}$
formalism described in \S 3.2.  The stellar masses for these systems and their 
integrated
stellar mass density evolution is described in detail in Papovich
et al. (2001), Dickinson et al. (2003) and Fontana et al. (2003).   
In Figure~12 mass functions are plotted as a function
of morphological type, while Figure~13 shows 
the mass density evolution, and the
fraction of mass in various morphological types, as a function of redshift.
These stellar mass densities are tabulated in Table~4 as a function of
redshift.

Figure~11 and 13 demonstrate that the stellar
mass density decreases as one goes to higher redshifts 
(Dickinson et al. 2003; Fontana et al. 2003; Rudnick et al. 2003; Drory
et al. 2004).  
Figure~11 \& 12 show that the decrease in
stellar masses is not due to any one type of
galaxy, but is the result of galaxies of all morphologies
growing in mass at lower redshifts.  Some
massive galaxies appear to have already formed between
$0.5 < z < 1.4$ (Figure~12) in the HDF-N.  Disks and ellipticals in
the HDF-N with stellar
masses M$_{*} > 10^{11.5}$ have a co-moving stellar mass 
density similar
to the total density of these objects at $z \sim 0$ (Cole et al. 2000).
This has been observed at higher
redshifts as well (Glazebrook et al. 2004; Franx et al. 2003) 
where some massive galaxies
appear to be already formed by $z \sim 1.5$.   While we do not see
massive galaxies in the HDF-S around $z \sim 1$ as we do in the HDF-N, we
do find massive HDF-S systems at $z = 2 - 3$.  This suggests that cosmic
variance is indeed a real issue when trying to understand the evolution of
galaxy properties using small HST fields.

Figures 12 \& 13 allow us to determine the types of galaxies contributing to
the stellar mass density at various redshifts.  We do not extrapolate our
stellar mass densities at the faint end to determine the total
stellar mass density, as done in Dickinson et al. (2003).  We ignore the 
faint galaxies we do not detect
as we are interested in the contribution of different morphological types
to the mass density, and the morphological types of these
fainter systems is unknown.   The total mass density shown in Figure~13 is
therefore an underestimate of the total (see Dickinson et al. 2003).  This
correction is however relatively minor.

In our lowest redshift range, $0.5 < z < 1.4$, the mass density is dominated
by ellipticals and spirals.  The fraction of mass in the peculiar phase is
very small, only $\sim 5$\% of the total in both the HDF-N and HDF-S.   
 The peculiar systems in
this redshift range also have low masses, nearly all with 
M$_{*} < 10^{10}$ \solm.  At higher redshifts, $z > 1.4$ there
is a dramatic change, such that morphologically identified 
ellipticals and disks contribute
a small fraction to the total stellar mass density while peculiars
contribute the bulk of it.  At the highest redshifts, $2.5 < z < 3$
peculiar galaxies contribute $>$ 60-80\% of the stellar mass density in
both fields. 
In Figure~14 we plot the redshift and morphological distribution of stellar 
masses out to $z \sim 4$. Figures~12 and 14 demonstrate that
at redshifts $z > 1.5$ the most massive systems are generally
peculiars, while at redshifts lower than $z \sim 1.5$ the most
massive systems are ellipticals. These trends can be further seen in 
Figure~13; 
just as in the luminosity density, the stellar mass densities of
spirals, ellipticals and peculiars are nearly equal at $z \sim 1.5$.

One major conclusion from our analysis of the number, luminosity and
stellar mass densities is that disks, ellipticals and peculiars have 
nearly equal co-moving number,
luminosity, and stellar mass densities at $z \sim 1.5$, and diverge at
higher and lower redshifts.  This is a smooth transition and several
factors indicate that peculiars at high 
redshift are possibly the ancestors of ellipticals
and components of disks found at lower redshift. 
High redshift peculiars are already
fairly massive, are undergoing star formation, and must exist in some form
at $z \sim 0$ and 1.  The reverse is also true - the ellipticals must have
formed stars at $z > 1.5$ since some have quite old stellar populations
(e.g., Stanford et al. 2004; Moustakas et al. 2004). We would thus likely 
see their progenitors
at higher redshifts and these are likely the peculiar systems.

\subsection{Mass to Light Ratios and Mass Assembly}

Figure~14 plots both individual stellar masses, and mass to light
ratios as a function of redshift for galaxies in the HDF-N and HDF-S.  As
already discussed, there is an increase in stellar masses both in the
aggregate, but also in terms of individual systems, out to about
$z \sim 1$.  Based on the right panel of Figure~14, which
shows the stellar mass to light ratios of the same galaxies plotted
in the left panel of Figure~14, we can draw some general conclusions.  
First, there are no galaxies with large M/L ratios until about $z \sim 2$.
This shows that in the HDF-N, star formation is very common in all
galaxies in our selection.  In fact, at $z > 2$ the vast majority of
all galaxies have mass to light ratios consistent with undergoing a
starbursts within the past 500 Myrs or less.  This may not be
a general result as there are more evolved high redshift galaxies
in the Hubble Deep Field South (Franx et al. 2003) and other fields
(Daddi et al. 2004). Some of these objects can be seen on Figure~14 as
systems with high M/L ratios at $z > 2$.
In general it appears that only by about $z \sim 1.5$ do
galaxies end starburst phases and begin to passively evolve.  Note
also that Figure~14a shows a selection effect such that only objects
with M $> 10^{10}$ \solm are visible at all redshifts regardless of their
mass to light ratios.

\subsection{Stellar Light Distributions: Interpreting Morphologies}

\subsubsection{CAS Values as Function of Type}

In addition to studying the apparent morphologies of galaxies in the two
Hubble Deep Fields we have  
measured their concentrations, asymmetries, and clumpiness (CAS) indices,
as well as a Petrosian and half light radii,  using the methodology outlined 
in Conselice (2003).
These quantify the stellar light distributions in the rest-frame B-band
for which we can determine a galaxy's form, shape, and size.

To understand how our eye-ball classifications fall within CAS space, we plot 
in Figure~15 concentration-asymmetry (C-A) diagrams for HDF-N and
HDF-S galaxies in four different redshift cuts.  Figure~16 shows the 
corresponding 
asymmetry-clumpiness (A-S) diagrams.  The lines
in the C-A diagram denote the differences between early and late types
(Bershady et al. 2000) as well as galaxies consistent with mergers
(Conselice et al. 2000a,b; Conselice 2003).  In the lowest redshift
range, $0.2 < z < 0.7$, the agreement is good between the Bershady et al. 
(2000)
classification of nearby galaxies in the C-A plane and the morphological
classifications done by eye.  There are a few disk galaxies
in the early type regime, but these are dominated by large bulges. There is
only one classified early-type found in the late-type area and the peculiars 
are located in the merger area.  This also holds at intermediate redshifts, 
$0.7 < z < 1.3$,
where the early types and late-types are well separated and all but
one of the galaxies with $A > 0.35$ is morphologically classified as
a peculiar.  This however beings to break down in the $1.3 < z < 2.0$ 
range, where galaxies classified as early types are found in the late type area
and the peculiars are found at nearly all A values, but typically are
towards the higher A range.  This is the case as well for
systems at $ z > 2.0$.  This demonstrates that the structural features
of galaxies, as well as their stellar populations can be inconsistent
with the physical interpretation of morphological types established
at $z \sim 0$ (e.g., Moustakas et al. 2004).

The A-S digram (Figure~16) is a perpendicular slice through CAS space to 
the A-C plane (Conselice 2003).   The S parameter correlates
with star formation, while the A parameter is affected by both star formation
and dynamical activity such as merging.  The solid line in Figure~16
shows the relationship between $A$ and $S$ found for nearby normal galaxies
in Conselice (2003).  The early type galaxies out to $z \sim 1.3$ 
have low clumpiness and asymmetry values, as is found for
ellipticals at $z \sim 0$ (Conselice 2003).  The disk galaxies at $z > 0.2$ 
have larger S and A values, but still follow the trend (solid line) 
between S and A found for nearby normal systems. 
The dashed lines are the 3 $\sigma$ scatter of normal
nearby galaxies from this relationship at $z \sim 0$.  
It was shown in Conselice 
(2003) that systems which deviate from this S-A relationship by more
than 3 $\sigma$ are consistent with undergoing major mergers.  This is 
analogous to 
the color-asymmetry relationship and its outlier method of identifying major 
mergers (Conselice et al. 2000a; Conselice et al. 2000b; Conselice 2003; 
Conselice et al. 2003a). 
Figure~16 shows that in every redshift bin, the galaxies that deviate
from the nearby S-A relationship are the peculiars.  There are a few
non-peculiars that also deviate, but these on closer inspection are systems
that have nearby companions and slight peculiarities, although many are 
still identifiable as spirals morphologically.

\subsubsection{Physical Interpretation of Galaxy Structure}

The CAS diagrams (Figures 15 \& 16)
can be compared directly with normal galaxies using Figure~12 from
Conselice (2003).    Making this comparison, there appears to be
several differences between the HDF and the Conselice (2003) normal
galaxy sample.  One obvious difference is that there
are few galaxies in the HDF-N or S that have high concentrations, which
massive ellipticals tend to have (Bershady et al. 2000; Conselice 2003).  
A large fraction of the HDF sample have concentration values that
are consistent with being disk-like or dwarf like (Conselice et al. 2002).  
The nearby comparison sample however contains only nearby very bright and 
large early  types.  The galaxies in the HDF are perhaps
more representative of the galaxy population than the bright selected
samples used in Conselice (2003).

We can argue that this is likely a real physical difference in the sense
that ellipticals 
at higher redshifts in the HDFs likely have lower stellar
and total masses than the most massive ellipticals in the nearby universe 
(see \S 3.3).
We argue this through the correlation between the stellar masses
of ellipticals derived by Papovich (2002) and the concentration
index. Previously in Conselice (2003) it was argued
that the concentration index is a good representation of the scale
of a galaxy in the sense that more massive galaxies have a higher
degree of light concentration. There is in fact, a good correlation 
between stellar mass and concentration at $z \sim 1$.  The
best fit between these two parameters can be represented by,

$${\rm log (M}_{*}) = (0.92\pm0.13) \times C + 7.65\pm0.43.$$

\noindent  Galaxies which are more concentrated are 
most massive, as argued also in Graham et al. (2001).  We cannot yet say
if this relationship evolves.   This correlation,
and the fact that there are not many bright M$^{*}$ galaxies in the HDF, 
confirms our interpretation that there
are ellipticals in the HDF at a range of masses and scales (see
also Stanford et al. 2004). This also confirms
that the relationship between scale and light concentrations (Graham
et al. 2001) holds out to at least $z \sim 1.5$ for evolved stellar 
populations.

\section{Discussion}

Ultimately, we are interested in connecting the high redshift
population with lower redshift galaxies, most notably if high redshift 
peculiars evolve into lower redshift 
ellipticals and bulges.   Since high redshift peculiars
are forming in massive starbursts (e.g., Steidel et al. 1999), they
could in principle evolve into massive galaxies.   
Color-asymmetry diagrams (Figure~17) give us some idea for which
types of galaxies are undergoing star formation and the triggering
mechanisms.  The solid and dashed
lines show the $z \sim 0$ relationship between (B-V) color and asymmetry
as well as its 3 $\sigma$ deviation.  At all redshift ranges, the peculiars
are the bluest and most asymmetric, on average.  The only exception is
in the $2.0 < z < 4.0$ redshift range where corrections to the asymmetry
values are substantial (Table~1). We argue from these figures
that many galaxies undergoing star formation are asymmetric,
and thus likely to be involved in major mergers (Conselice et al. 2003).  
That is, star formation appears
to be merger induced in a significant fraction of peculiar galaxies (Mobasher
et al. 2004).

What is the fate of these peculiar galaxies which appear to be involved
in major mergers?  There are other clues that suggest they must
become the Hubble types at lower redshifts.  The first
is that the co-moving stellar mass density of peculiars declines with
redshift (Figure~13; Table~4).  Although the mass density appears to remain
constant at $z > 1$ for the peculiars, these systems are undergoing massive 
star formation
and thus should increase in stellar mass density over time unless they
evolve morphologically into different types.    
 The mass in peculiars must go somewhere, and
given their stellar masses and the characteristics of their halos (Giavaliso
et al. 1998) they are likely in some form the progenitors of lower redshift
normal galaxies.  Another clue is that these systems are merging and 
starbursting.
By integrating the amount of stellar mass added to these galaxies over
time through mergers and starbursts, their stellar masses could become as large
as a modern M$^{*}$ galaxy (Papovich et al. 2001; Conselice 2004, 
in preparation).  The most
massive galaxies appear to be nearly formed by $z \sim 1$, consistent
with the drop in major merger rates at $z < 1.5$ (Conselice et al. 2003).
It is also not likely that we are missing a population of dusty galaxies,
as sub-mm sources are detectable in the optical, within our limits, and also
have peculiar morphologies (Conselice, Chapman \& Windhorst 2003; Chapman
et al. 2003).

Major mergers and the star formation it induces is therefore likely 
responsible for the luminosity density at $z > 2$ and the build up of
stellar mass to $z \sim 1.5$. As argued in Conselice (2003) and Conselice
et al. (2003a) when using the rest-frame B-band light from a galaxy only
major mergers produce a high asymmetry, these are thus unlikely to
be systems undergoing minor mergers.   Star formation, even massive
amounts, does not necessarily produce highly asymmetric galaxies either 
(Conselice et al. 2000a).   In any case, only about 50 - 75\% of all
stellar mass is assembled by this redshift.  There must be other methods
whereby galaxies are forming stars at lower redshift.  The methods
must be such that they do not add significant mass to the most massive
galaxies (\S 3.3), nor can they significantly change the morphologies of 
these galaxies (\S 3.1), thus major mergers are unlikely the cause.  

We cannot determine with the present data what the cause of the increase in 
stellar
mass at $z < 1.5$ is produced by.  Minor mergers and accretion of
intergalactic gas are the two major possibilities. Recently, the minor
merger history was traced using deep near infrared imaging out to $z \sim 1$
by Bundy et al. (2004).  Bundy et al. (2004) found that the pair fractions
of galaxies increases with redshift, but not as strongly as
pair fractions detected in optical surveys.  This implies that the mass to
light ratios of the accreted galaxies are low and that they are undergoing
star formation. Bundy et al. (2004) calculate that the amount of stellar mass
in these satellites is roughly equal to the increase seen between
$z \sim 1$ to 0.  However, these masses are accounted for within the
stellar mass density calculations (Dickinson et al. 2003).  Since
these satellites appear to be undergoing interaction induced star formation
this increase might be enough to account for the additional stellar mass
accreted in galaxies at $z < 1$.  Star formation histories and rates
for these satellites and their hosts will need to be measured to address
this possibility.

\section{Summary}

In this paper we present an analysis of rest-frame B-band luminosity,
morphology, and stellar mass as a function of redshifts out to $z \sim 3$
using the Hubble Deep Field North and South. The aim of this study is
to determine the evolution of galaxy structure and how it relates to 
the assembly of stellar mass in galaxies.  Our major conclusions are:

1. Through a visual analysis of the morphologies of galaxies in the HDF-N
and HDF-S we conclude that there are only a few classifiable ellipticals
or disks at redshift $z > 1.5$.  This drop is abrupt and is robust to
changes in S/N, resolution and morphological k-corrections at $>$ 4 $\sigma$
significance based on simulations.  There also appears in the HDF-N and HDF-S  
a gradual decline in the co-moving density of spirals and ellipticals
as a function of redshift.

2. The decline in the number of ellipticals and disks at $z > 1.5$ mirrors
the rise in the number of peculiars at higher redshifts.  At $z \sim 2.5$
approximately 60\% at I $< 27$ of all galaxies are peculiars.  A significant
fraction of galaxies at higher redshift are unclassifiable, with $\sim$ 50\%
at $z \sim 3.5$ too faint for a reliable classification.

3. We investigate the luminosity function of galaxies in the HDF-N and
HDF-S and compute the luminosity density as a function of morphological
type. We find, as others, that the luminosity density is dominated by
disks and ellipticals at $z < 1.5$ and by peculiar galaxies at $z > 1.5$.
We find that at $z \sim 0.4$ $\sim 90$\% of the luminosity density in
the HDF fields is in the form of disks and ellipticals, while at $z \sim 3$, 
60-80\% of the luminosity density originates from peculiar galaxies.

4. The change in relative number and luminosity densities of peculiars and 
normal galaxies is mirrored in the evolution of stellar mass as a function of
morphological type.  At redshifts $z \sim 1$, the most massive
galaxies are ellipticals, with disks generally of lower mass, and peculiars
having the lowest masses.  At $z > 2$, the most massive galaxies are
peculiars, although all have stellar masses less than M$^{*}$.  When
examining the density of stellar mass as a function of morphology we find
that at high redshift, $z > 2$, $\sim 50-80$\% of stellar mass is
in peculiars.  At $z < 1$, this fraction drops close to zero with a 
corresponding rise in the fraction and density of mass in ellipticals and
disks.  Importantly, the total absolute stellar mass density
of galaxies in a peculiar phase slightly declines, suggesting that 
peculiar galaxies have evolved into modern galaxies.

5. We compare the stellar and star forming properties of galaxies in the HDF-N
with their structural features using the CAS morphological system.  We find
that out to $z \sim 3$ eye-ball estimates of morphology agree well
with the automated CAS approach. That is, ellipticals out to redshifts
$z \sim 1.2$ are symmetric,
concentrated and smooth, while disks are less concentrated and more asymmetric
and clumpy, similar to what is found in the nearby universe (Conselice 2003).
The peculiar galaxies are identifiable as mergers based in the
CAS systems, having large rest-frame B-band asymmetries ($A_{\rm B}$).  These
asymmetric peculiars are also the bluest and are undergoing the most rapid
unobscured star formation.  

We conclude that  that some massive galaxies originate
from peculiars at high redshift.  These peculiar galaxies, particularly
the brighter ones, are undergoing major mergers (Conselice et al. 2003).
Major mergers are however not likely
to produce all the stellar mass in galaxies at $z < 1$ as a significant
fraction of stellar mass forms after $z \sim 1$ (Dickinson et al. 2003) and
there are few mergers during this time (Conselice et al. 2003).
It is possible that this addition comes in the form of minor mergers
(Bundy et al. 2004).  Future studies with the GOODS survey (Giavalisco
et al. 2004) will reveal in more detail this evolution and its
relationship to other physical properties.

We thank Mark Dickinson for his many contributions to the HDF NICMOS imaging
and Richard Ellis for comments on this paper.  We also thank the
anonymous referee for several valuable suggestions that have improved this 
paper. This work was supported by a NSF Astronomy and 
Astrophysics Fellowship, and by NASA HST grant HST-AR-09533.04-A.  We also
acknowledge the support of a Caltech Summer Undergraduate Research Fellowship
(SURF) to JAB.

\newpage

\begin{figure}
\plotfiddle{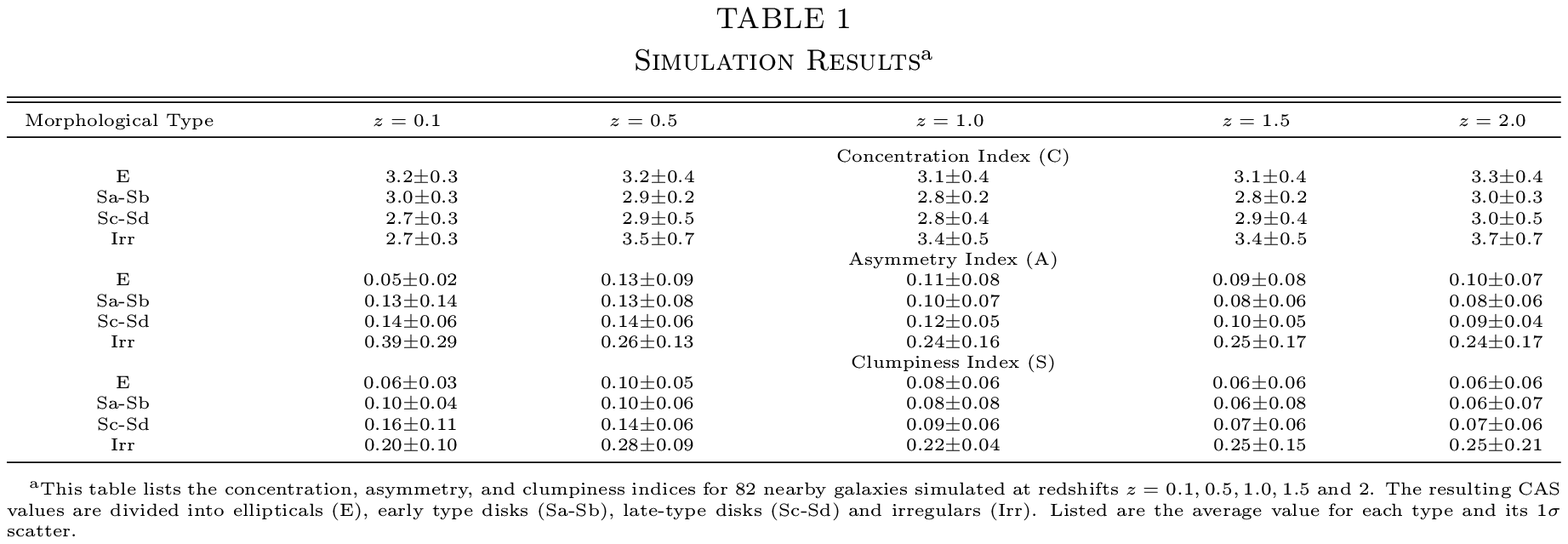}{6.0in}{0}{100}{100}{-310}{-170}
\vspace{0.7in}
\end{figure}
\begin{figure}
\plotfiddle{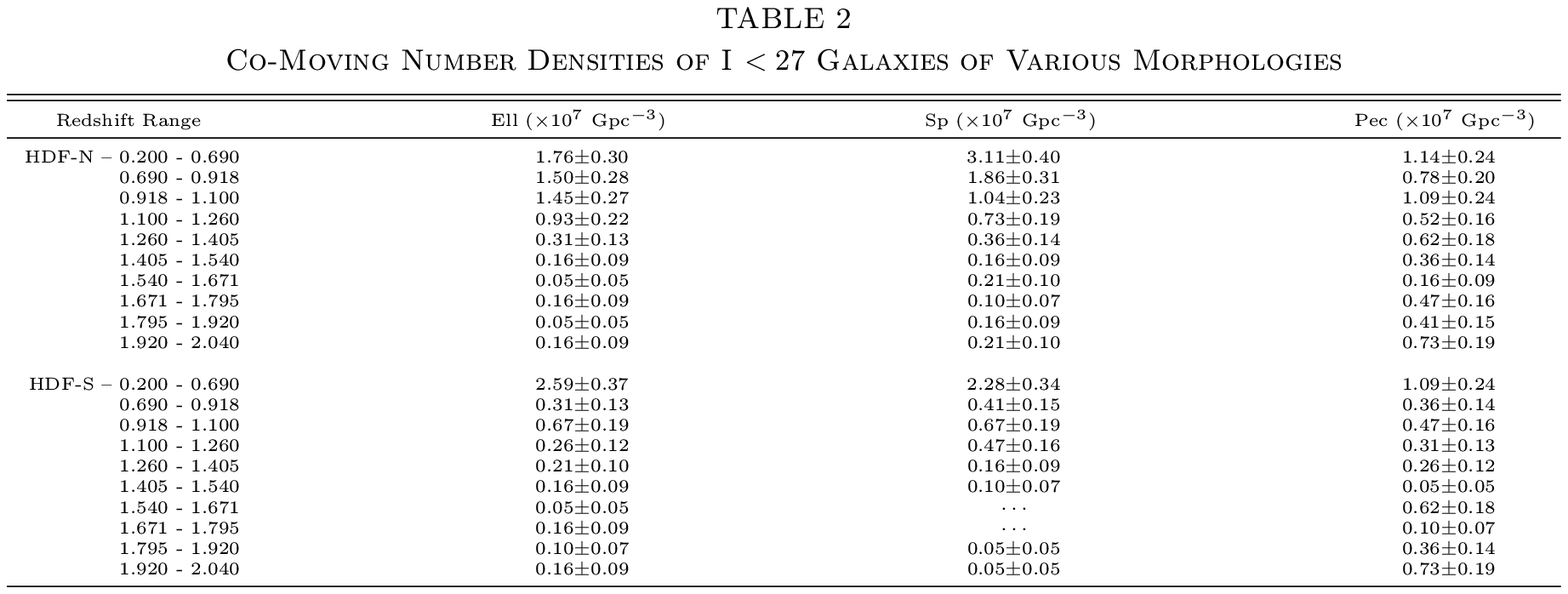}{6.0in}{0}{100}{100}{-310}{-170}
\vspace{0.7in}
\end{figure}
\begin{figure}
\plotfiddle{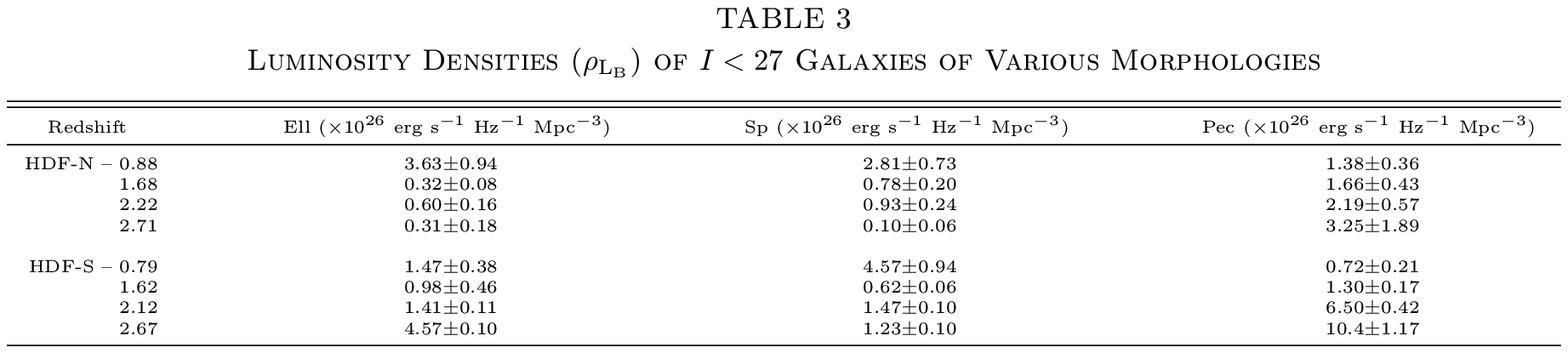}{6.0in}{0}{100}{100}{-310}{-170}
\vspace{0.7in}
\end{figure}
\begin{figure}
\plotfiddle{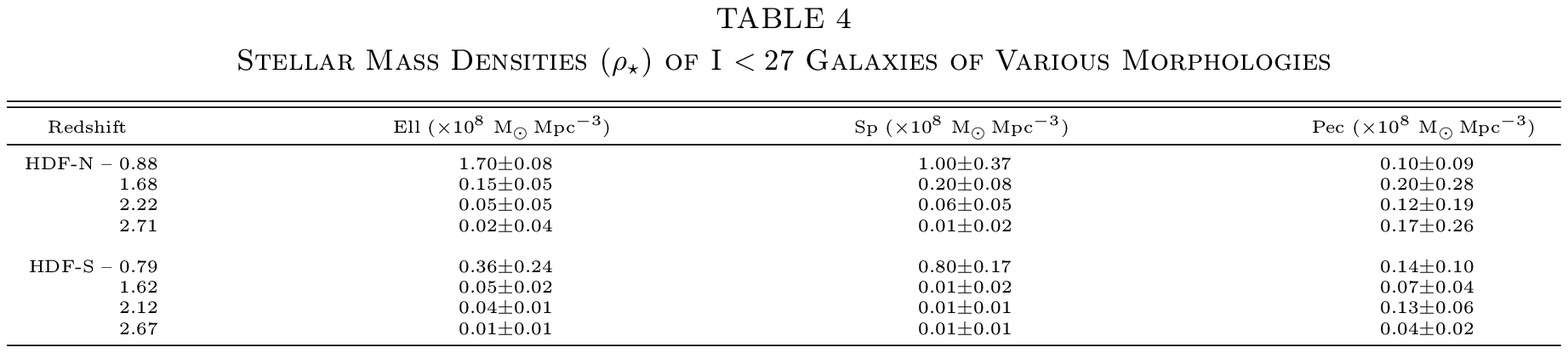}{6.0in}{0}{100}{100}{-310}{-170}
\vspace{0.7in}
\end{figure}

\setcounter{figure}{0}

\begin{figure}
\plotfiddle{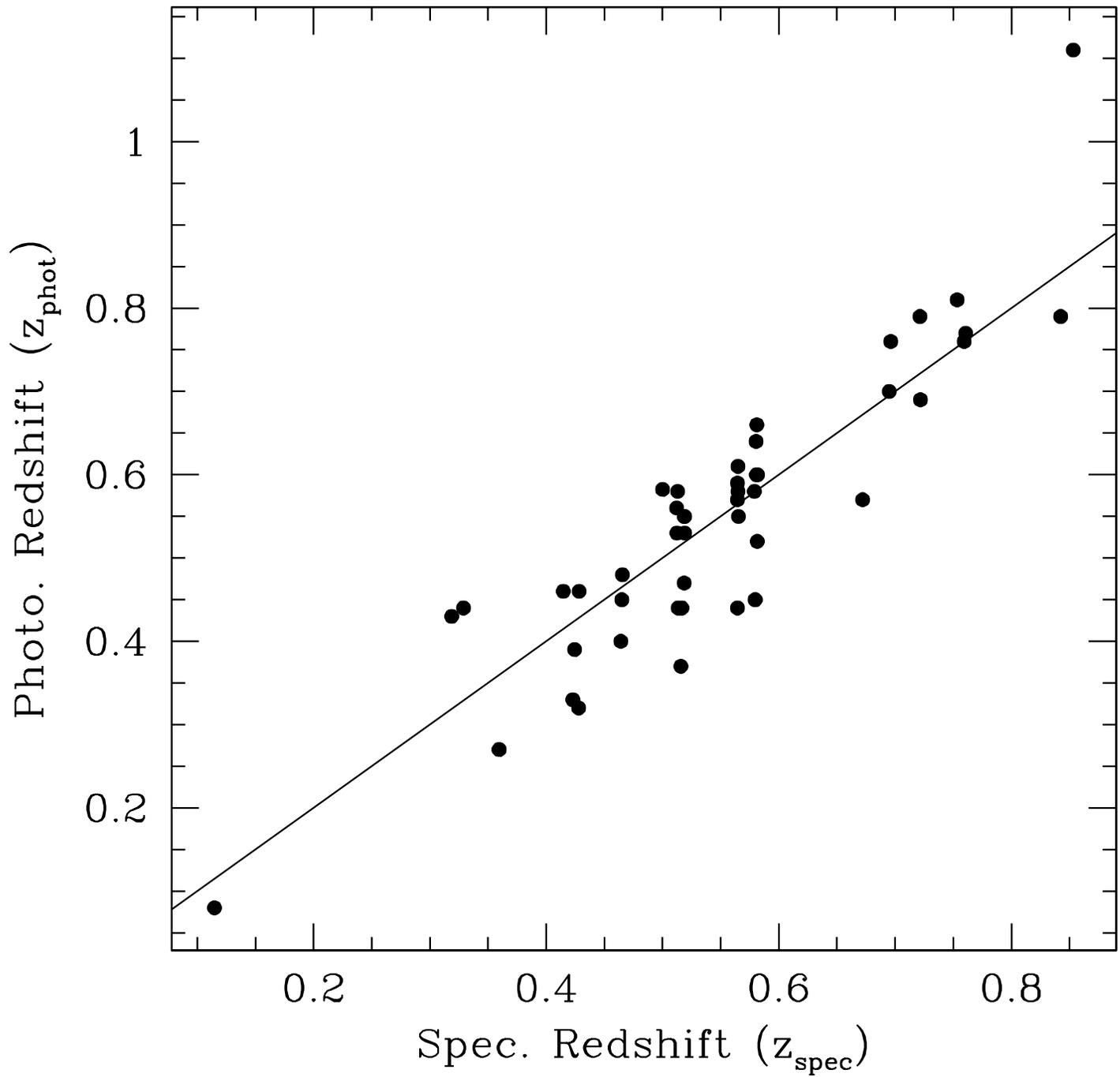}{6.0in}{0}{100}{100}{-310}{-170}
\vspace{0.7in}
\caption{The comparison between spectroscopic redshifts ($z_{\rm spec}$) 
measured
by Sawicki \& Mallen-Ornelas (2003) and the photometric redshifts ($z_{\rm 
phot}$) from Lanzetta et al. (2000).  The solid line shows the
relationship  $z_{\rm spec}$  = $z_{\rm phot}$.}
\end{figure}

\begin{figure}
\plotfiddle{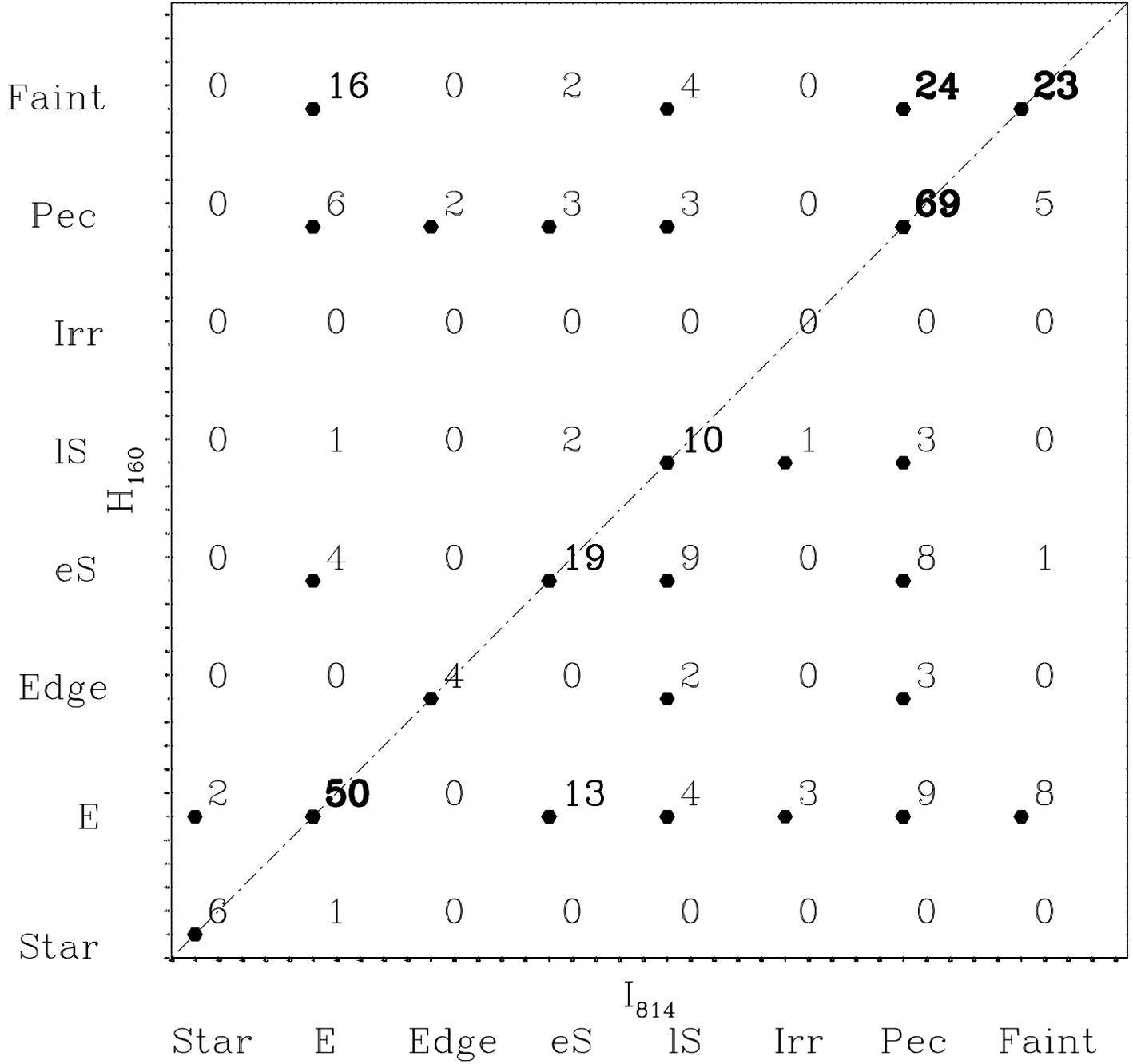}{6.0in}{0}{100}{100}{-310}{-170}
\vspace{0.7in}
\caption{The relationship between classifications in the I$_{814}$ band
and the H$_{160}$ band.  The number by each point lists the number of
galaxies that are classified in  I$_{814}$ and H$_{160}$ at that point.
The diagonal dot-dashed line traces covers the points
where classifications are the same in both bands.  Larger numbers
have a type which is weighted heavier.}
\end{figure}

\begin{figure}
\plotfiddle{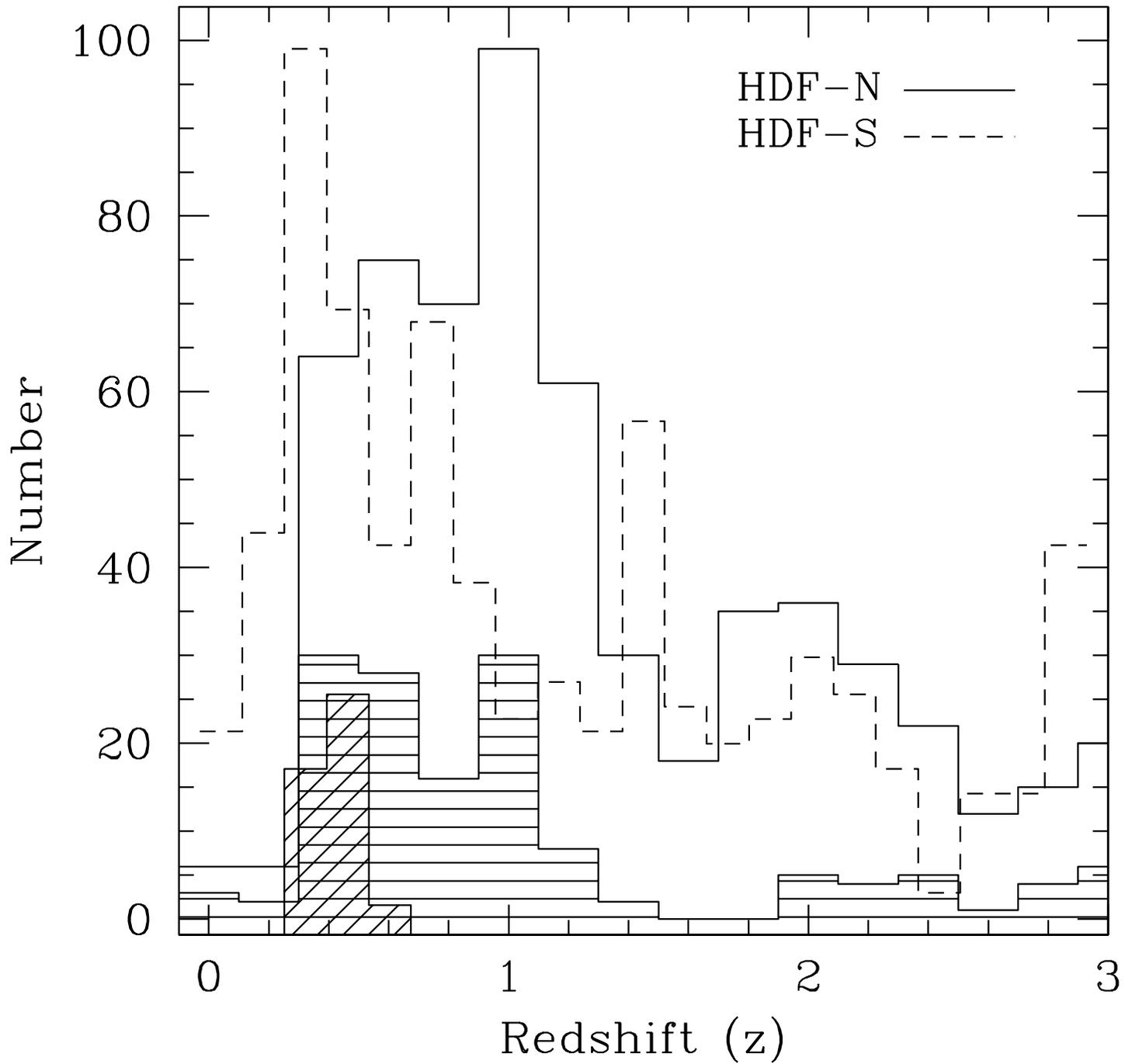}{6.0in}{0}{100}{100}{-310}{-170}
\vspace{0.7in}
\caption{Histogram of redshifts (both spectroscopic and photometric) 
for galaxies brighter than I$_{814} = 27$ in the HDF-N (solid) and HDF-S 
(dashed).  The horizontally shaded histogram displays the spectroscopic
redshifts for the HDF-N, and the diagonally shaded histogram is for
the spectroscopic redshifts in the HDF-S}
\end{figure}

\begin{figure}
\plotfiddle{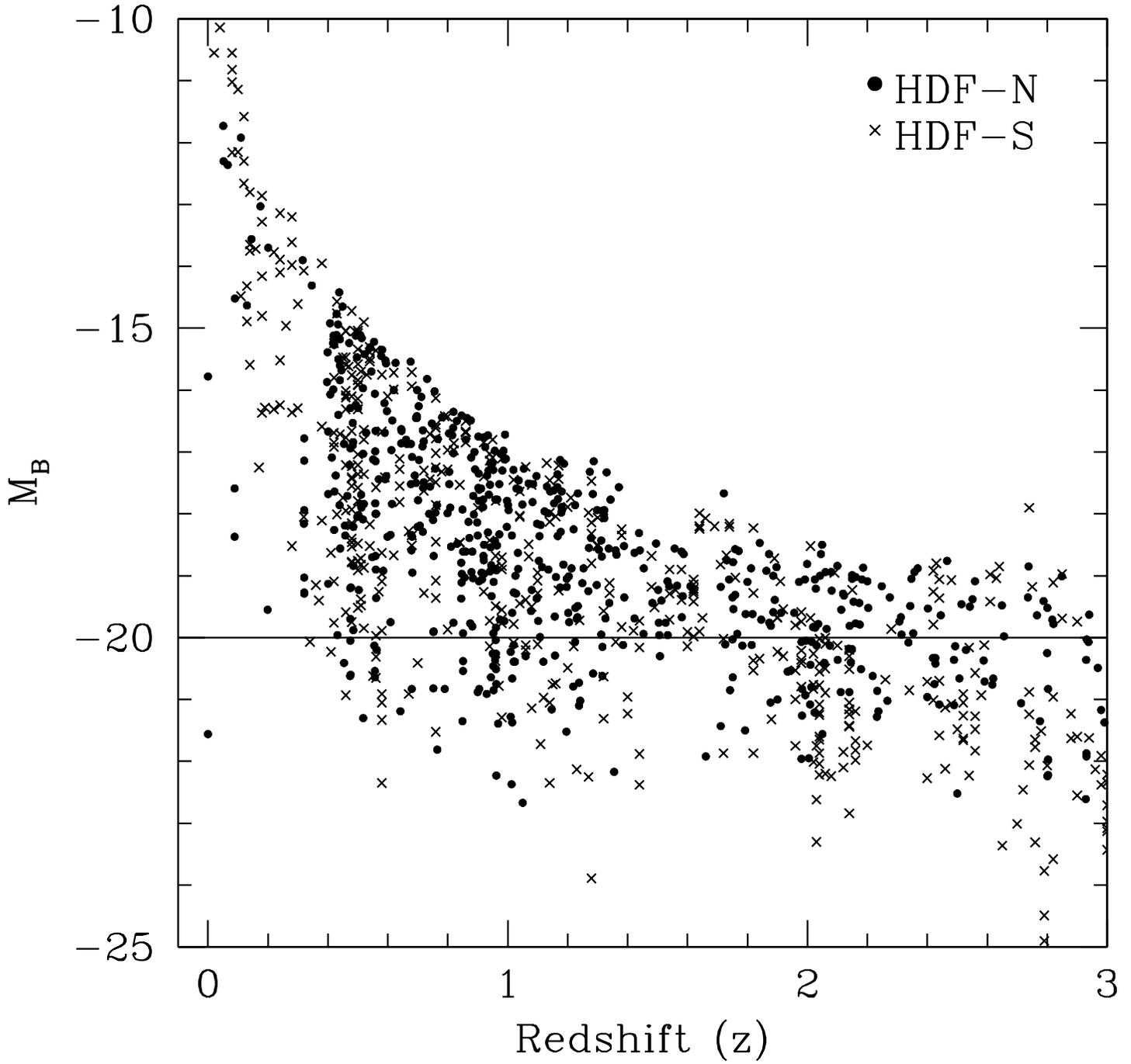}{6.0in}{0}{100}{100}{-310}{-170}
\vspace{0.7in}
\caption{The distribution of absolute magnitudes (M$_{\rm B}$) vs. redshift
for galaxies in the HDF-N (solid) and HDF-S (crosses).  The horizontal
dashed line shows the M$_{\rm B} = -20$ limit that we use for some analyses
in this paper.}
\end{figure}

\begin{figure}
\plotfiddle{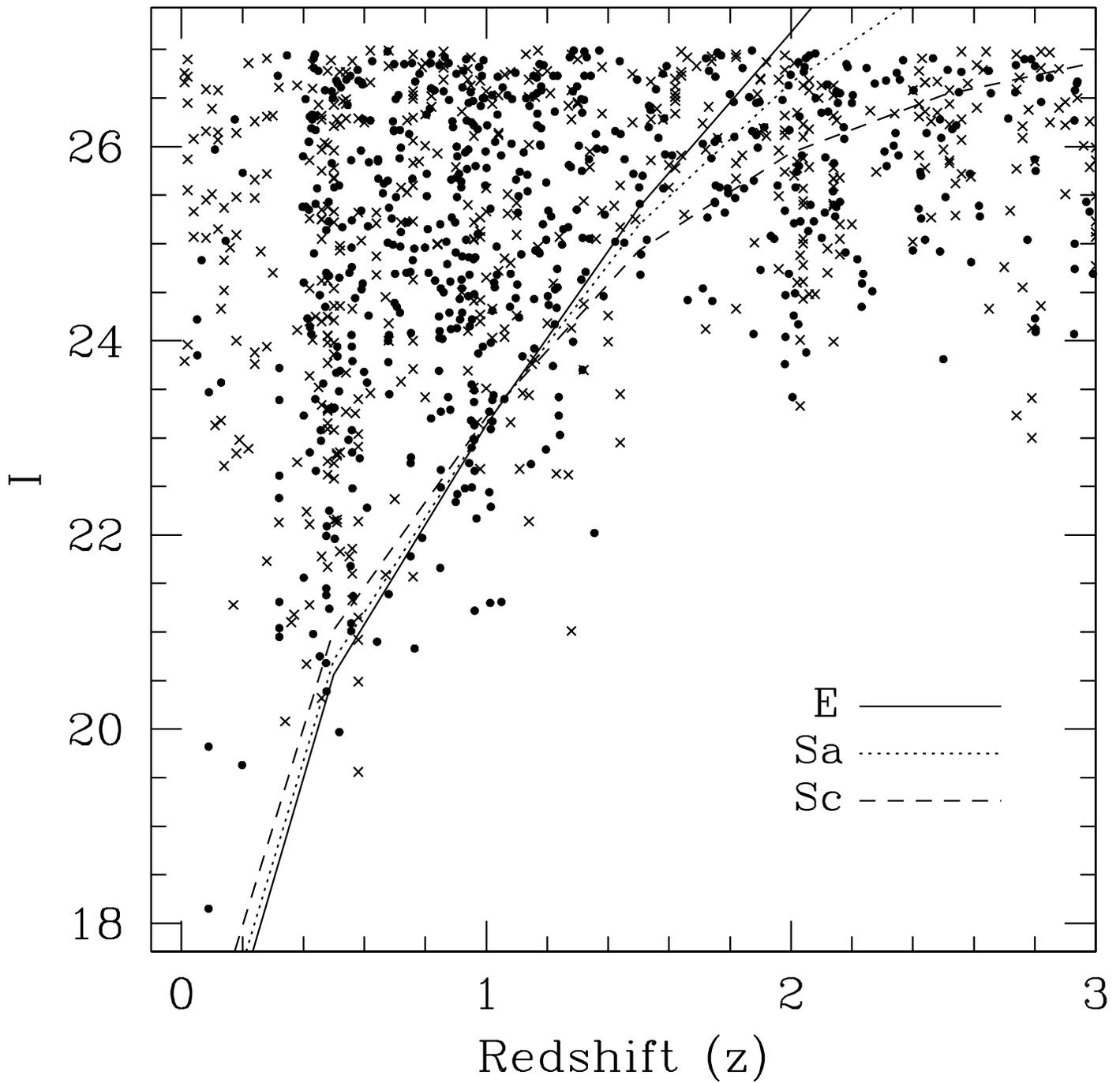}{6.0in}{0}{100}{100}{-310}{-170}
\vspace{0.7in}
\caption{The I$_{814}$ distribution for galaxies in the HDF-N (solid) and 
HDF-S (crosses) with I $< 27$.
The lines are k-corrected apparent I-band magnitudes for M$_{\rm B} = -20$
ellipticals and early/late spirals at these redshifts.}
\end{figure}

\begin{figure}
\plotfiddle{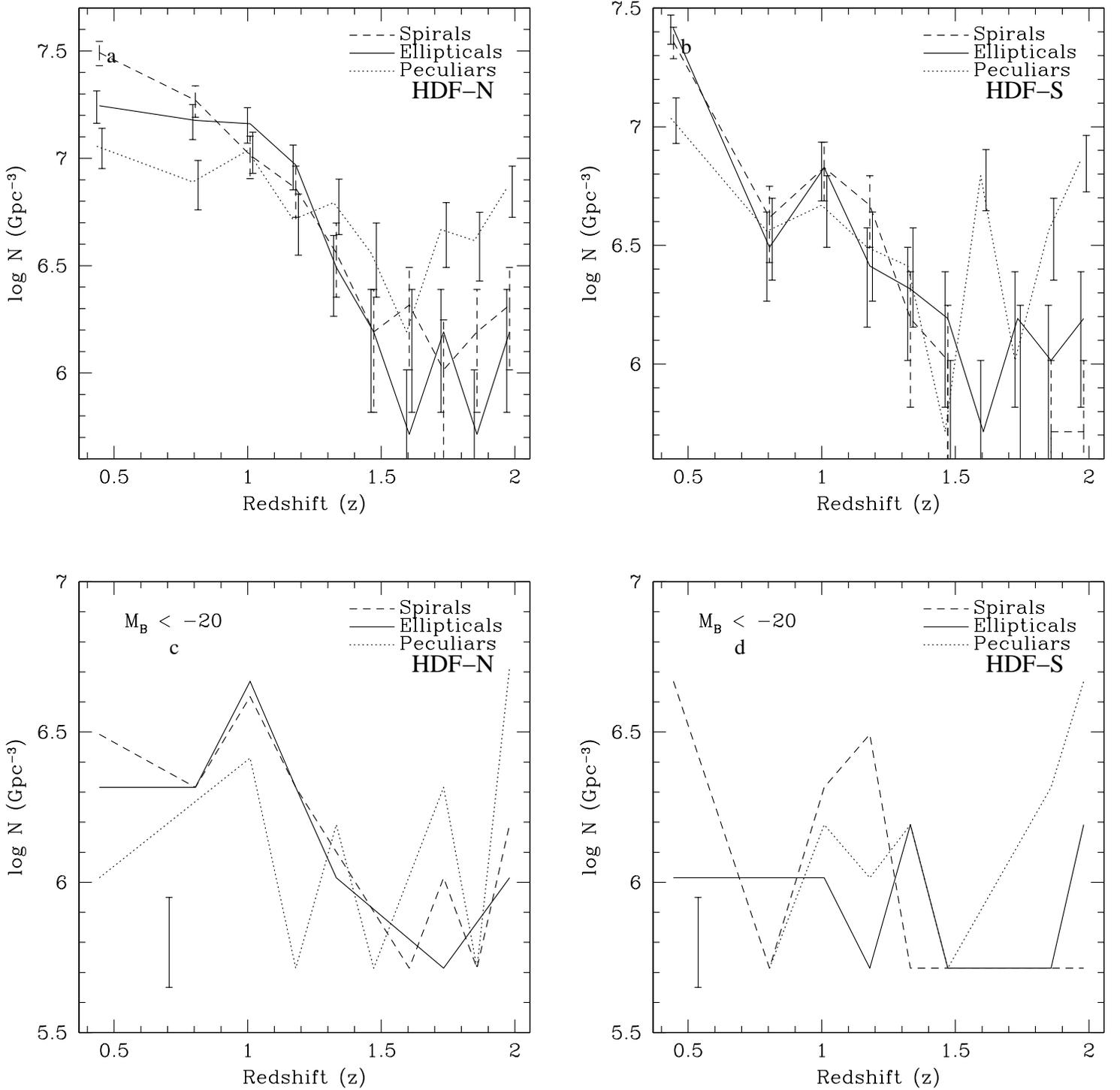}{6.0in}{0}{100}{100}{-310}{-170}
\vspace{0.7in}
\caption{The co-moving number density evolution of ellipticals, spirals and
peculiars as a function of redshift in the HDF-N and HDF-S.  The upper panel 
is for the total
sample of galaxies with $I < 27$ and the bottom panel is the co-moving
number densities for those galaxies which are brighter than M$_{\rm B} = -20$.}
\end{figure}

\begin{figure}
\plotfiddle{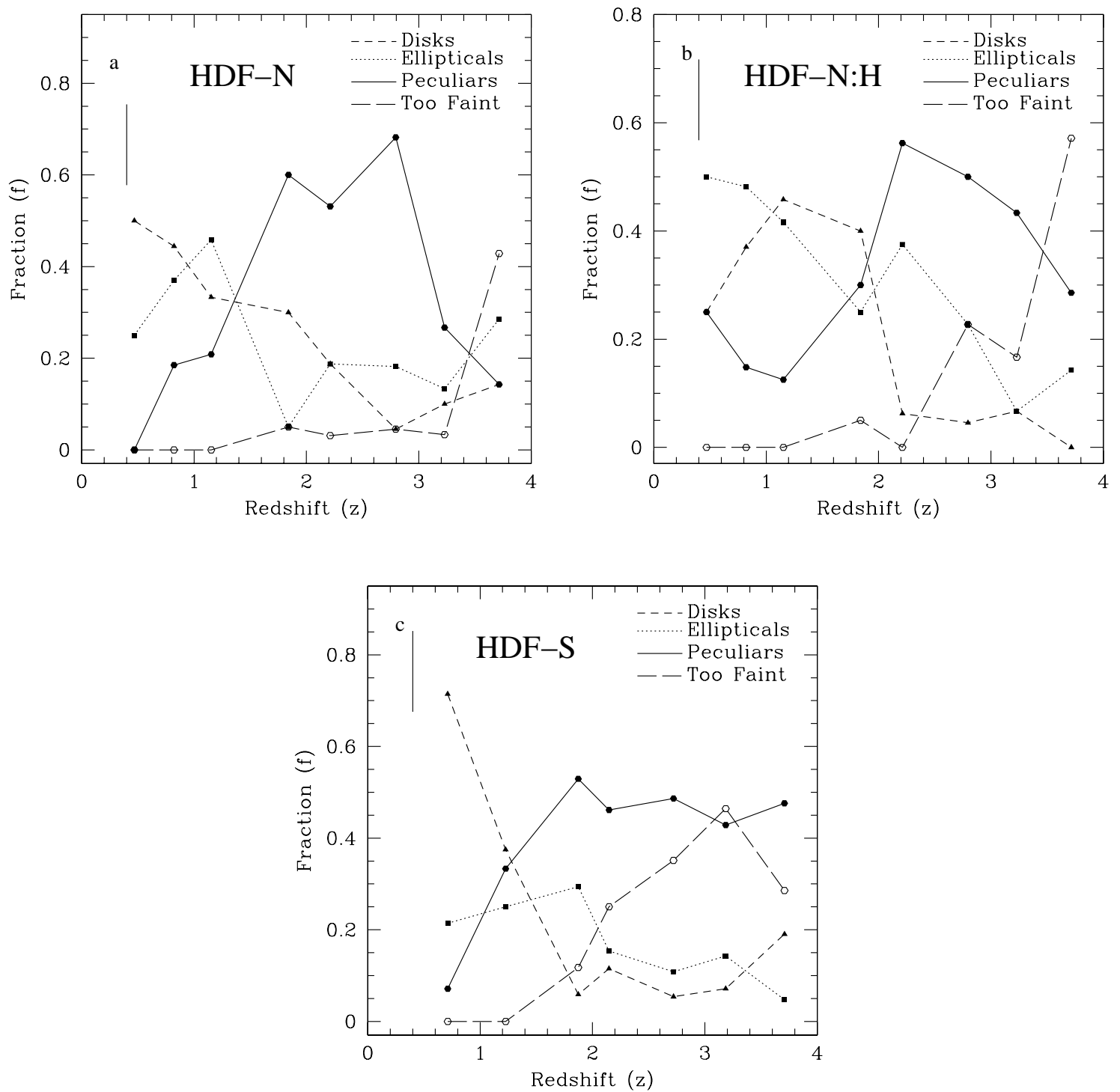}{6.0in}{0}{100}{100}{-310}{-170}
\vspace{0.7in}
\caption{Evolution in relative fractions of different galaxy types
as a function of redshift for classifications in the I$_{814}$
and H$_{160}$ band images of the HDF-N (a and b) and for the I$_{814}$ image
of the HDF-S (c).  The vertical solid line on each plots gives the average
error for these fractions.  Types plotted on this are: disks (short dashed), 
ellipticals (dotted),
peculiars (solid) and galaxies which are too faint for a classification 
(long dashed).}
\end{figure}

\begin{figure}
\plotfiddle{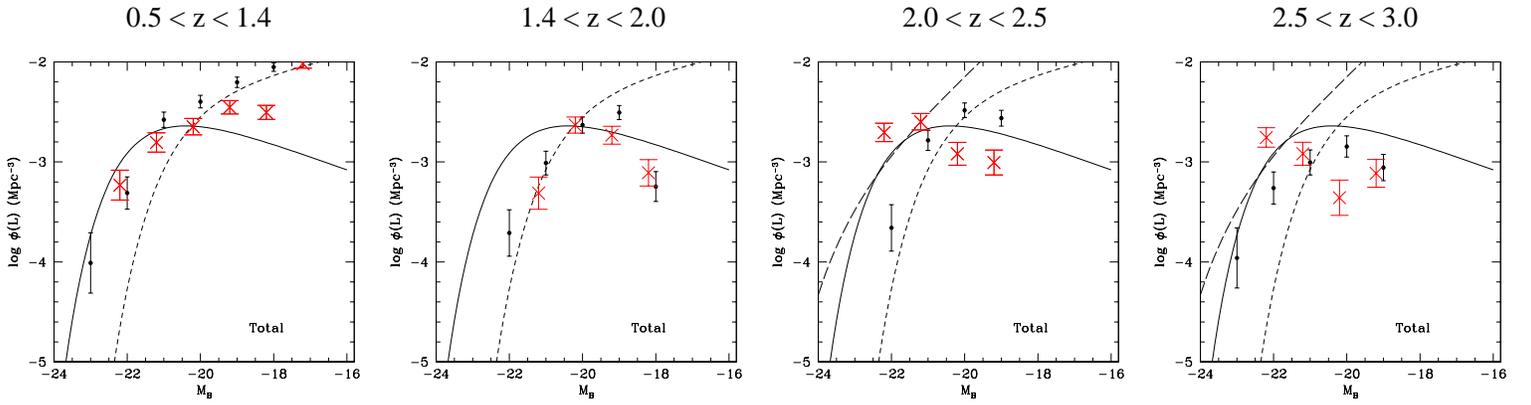}{6.0in}{0}{100}{100}{-310}{-170}
\vspace{0.7in}
\caption{The total luminosity function for galaxies in the HDF-N (solid
black points) and the HDF-S (large crosses) plotted as a function of
redshift for the I$_{814} < 27$ sample.  The dashed line is the 2dF B-band 
luminosity function (Norberg
et al. 2002), the solid line is the CFRS $z \sim 1$ luminosity function
(Lilly et al. 1995) and the dashed line plotted for panels at $z > 2$
is the rest-frame B-band Lyman-break galaxy luminosity function  
(Shapley et al. 2001).}
\end{figure}

\begin{figure}
\plotfiddle{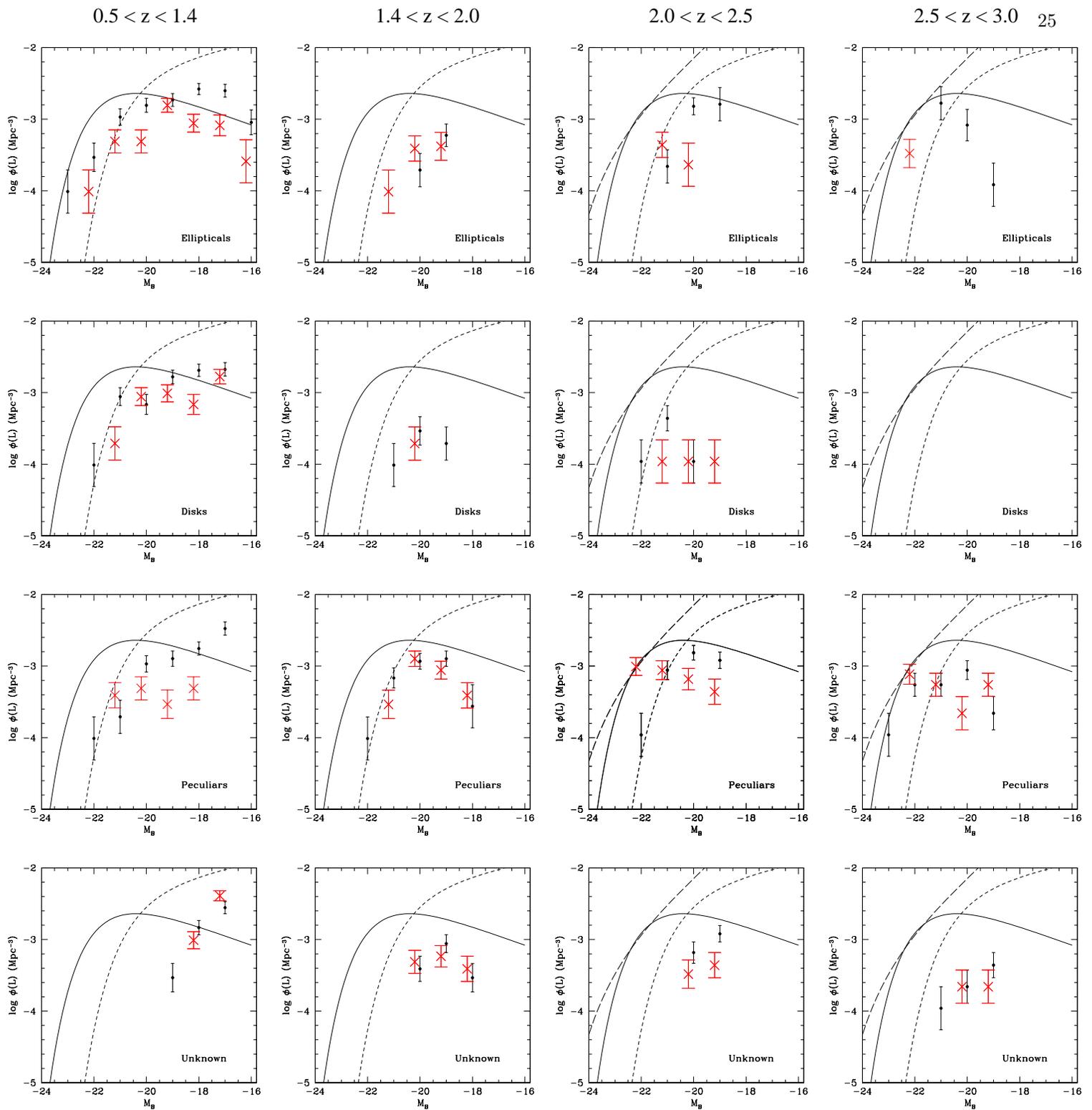}{6.0in}{0}{100}{100}{-310}{-170}
\vspace{1.7in}
\caption{The luminosity function as function of morphological type and
redshift.  The points and lines are the same as in Figure~8.  }
\end{figure}
\clearpage

\begin{figure}
\plotfiddle{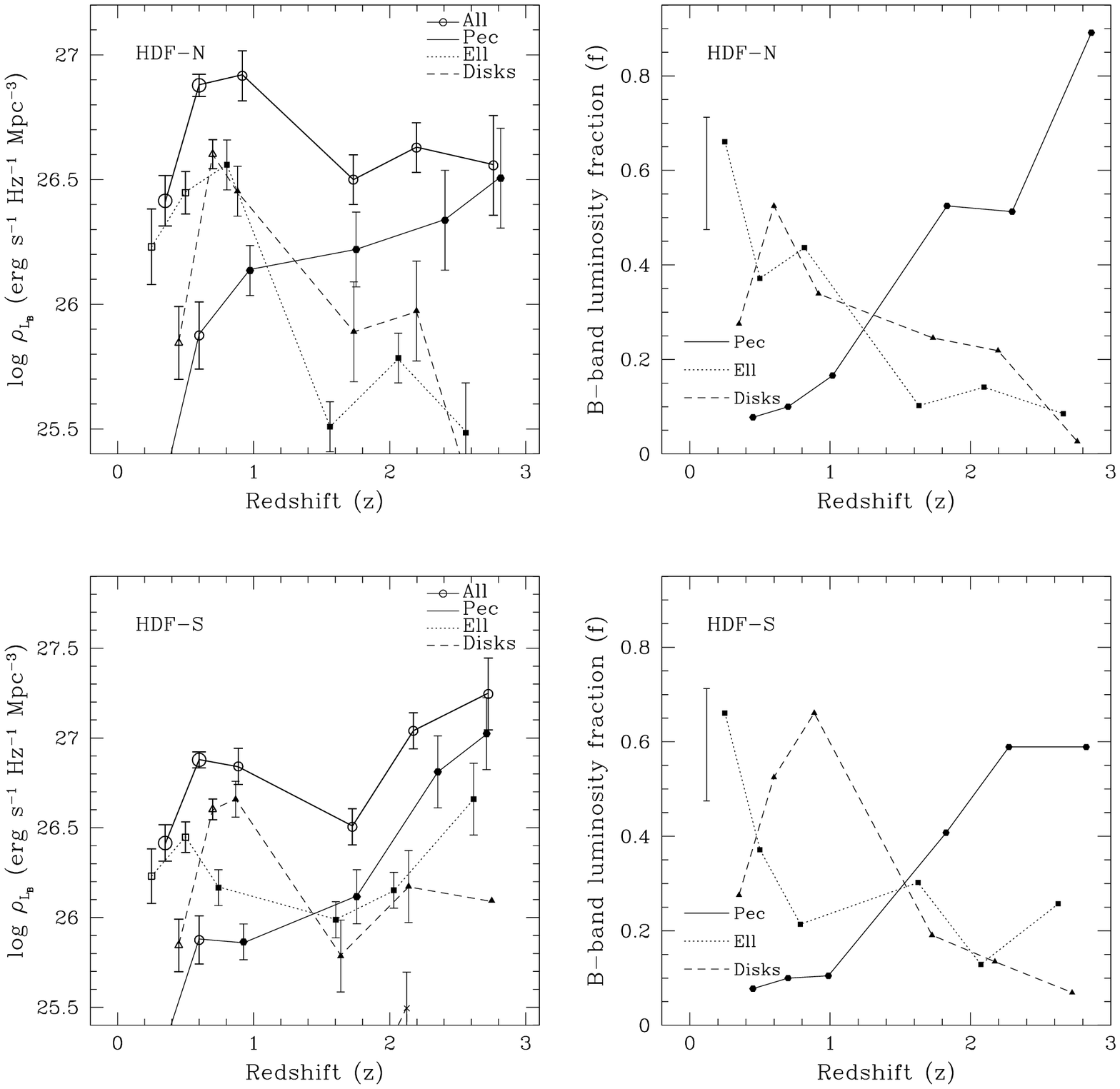}{6.0in}{0}{100}{100}{-310}{-170}
\vspace{0.7in}
\caption{The luminosity density and luminosity fraction as a function of 
morphological type and redshift.  The upper solid line and open circles
shows the total B-band luminosity density in the the HDF-N and HDF-S.  The
solid line and solid circles show the luminosity density evolution for
peculiar galaxies, while the triangles and dashed line and the boxes and 
dotted lines are the luminosity densities for disks and ellipticals, 
respectively.
The larger open symbols at $z < 1$ for the total and morphologically
divided luminosity densities are taken from Brinchmann \& Ellis (2000).}
\end{figure}

\newpage

\begin{figure}
\plotfiddle{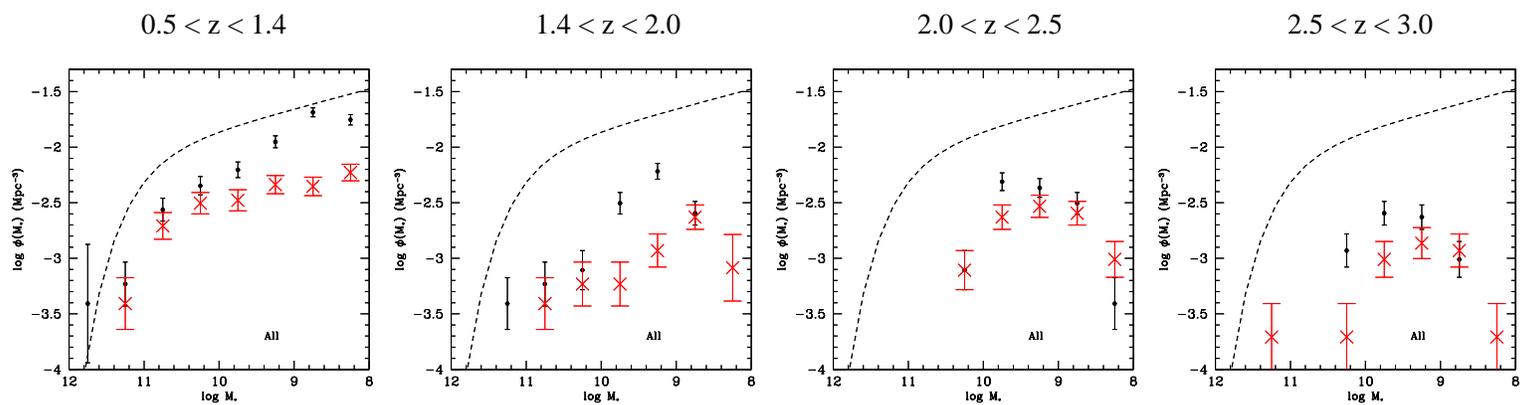}{6.0in}{0}{100}{100}{-310}{-170}
\vspace{0.7in}
\caption{The total mass function as a function of redshift for the HDF-N 
(solid black points) and the HDF-S (large crosses) plotted in four
different redshift bins  for the I$_{814} < 27$ sample. 
The dashed line shows the $z \sim 0$ mass function from 2dF+2MASS observations
(Cole et al. 2001). }
\end{figure}

\begin{figure}
\plotfiddle{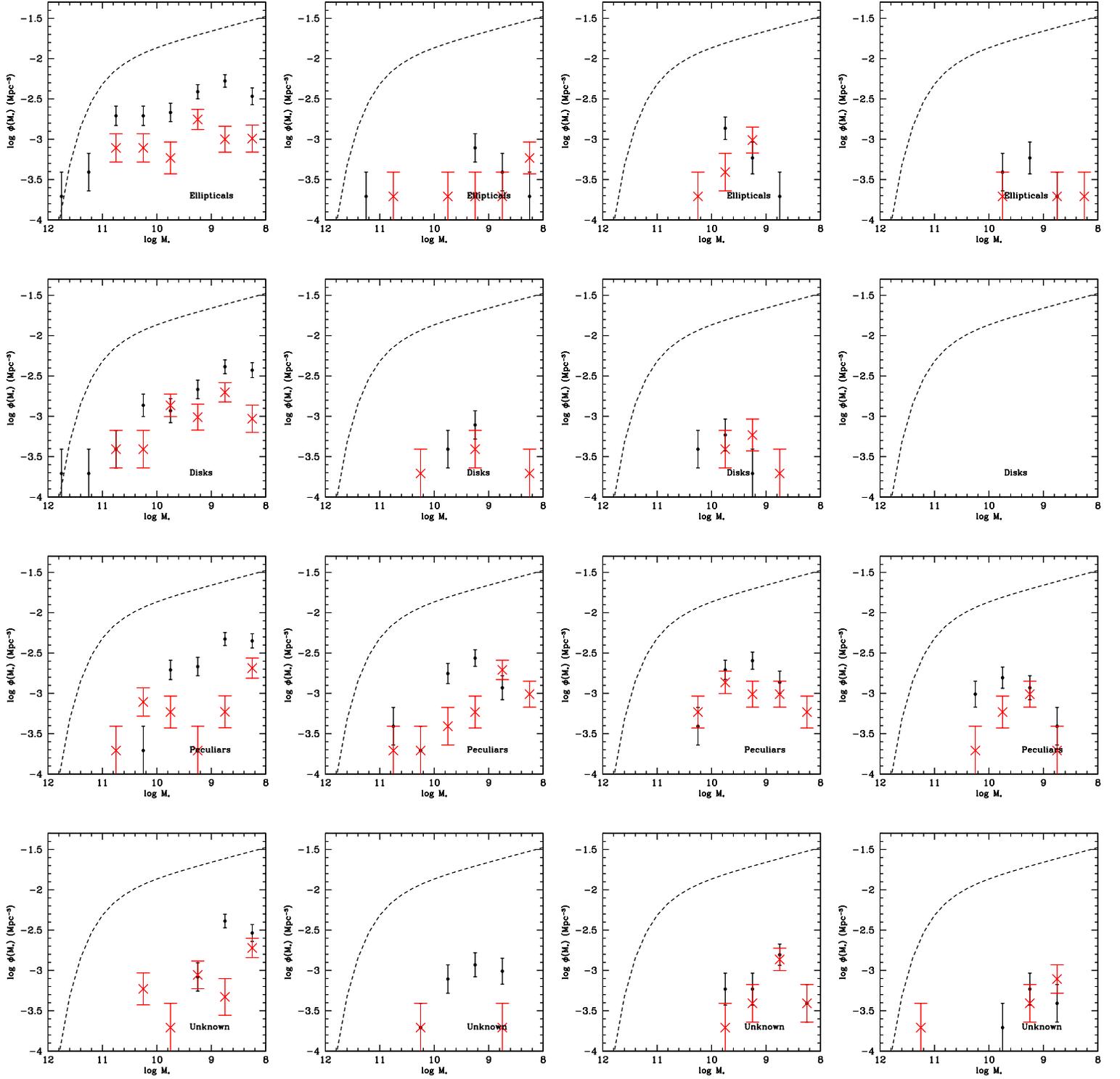}{6.0in}{0}{100}{100}{-310}{-170}
\vspace{1.2in}
\caption{Mass functions as a function of morphological type.  The dashed
line is the 2dF+2MASS mass function (Cole et al. 2001). Lines and symbols
are the same as in Figure~11.}
\end{figure}

\clearpage
\begin{figure}
\plotfiddle{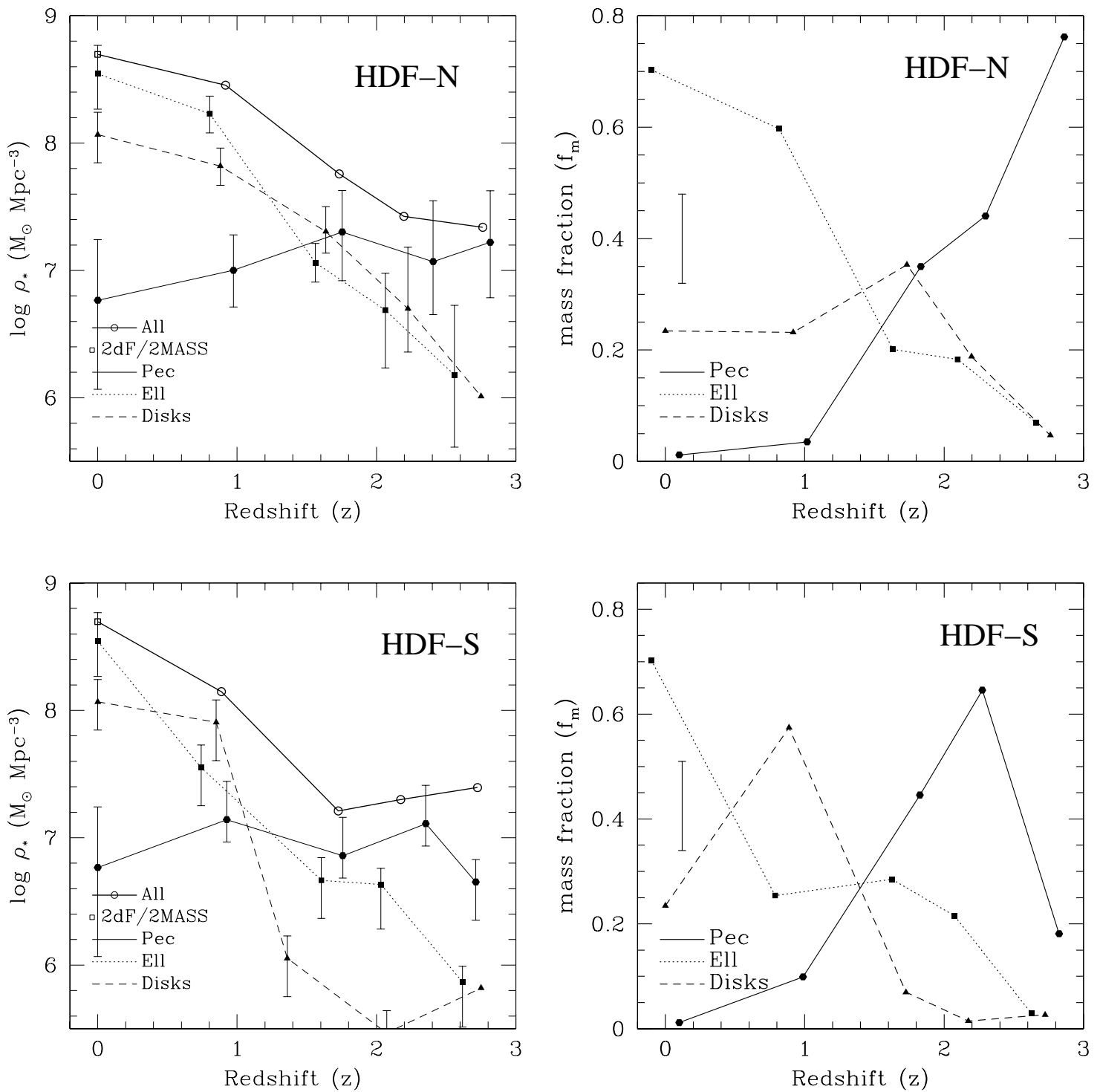}{6.0in}{0}{100}{100}{-310}{-170}
\vspace{0.7in}
\caption{The evolution and fraction of stellar mass as a function of 
morphology out
to $z \sim 3$.  As in Figure~10, the solid upper line is the total mass 
density as a function
of redshift with the $z \sim 0$ point from the 2MASS data of Cole et al. 
(2001).  The
$z \sim 0$ points for the elliptical and disk stellar masses are taken
from Fukugita et al. (1998). }
\end{figure}

\clearpage
\begin{figure}
\plotfiddle{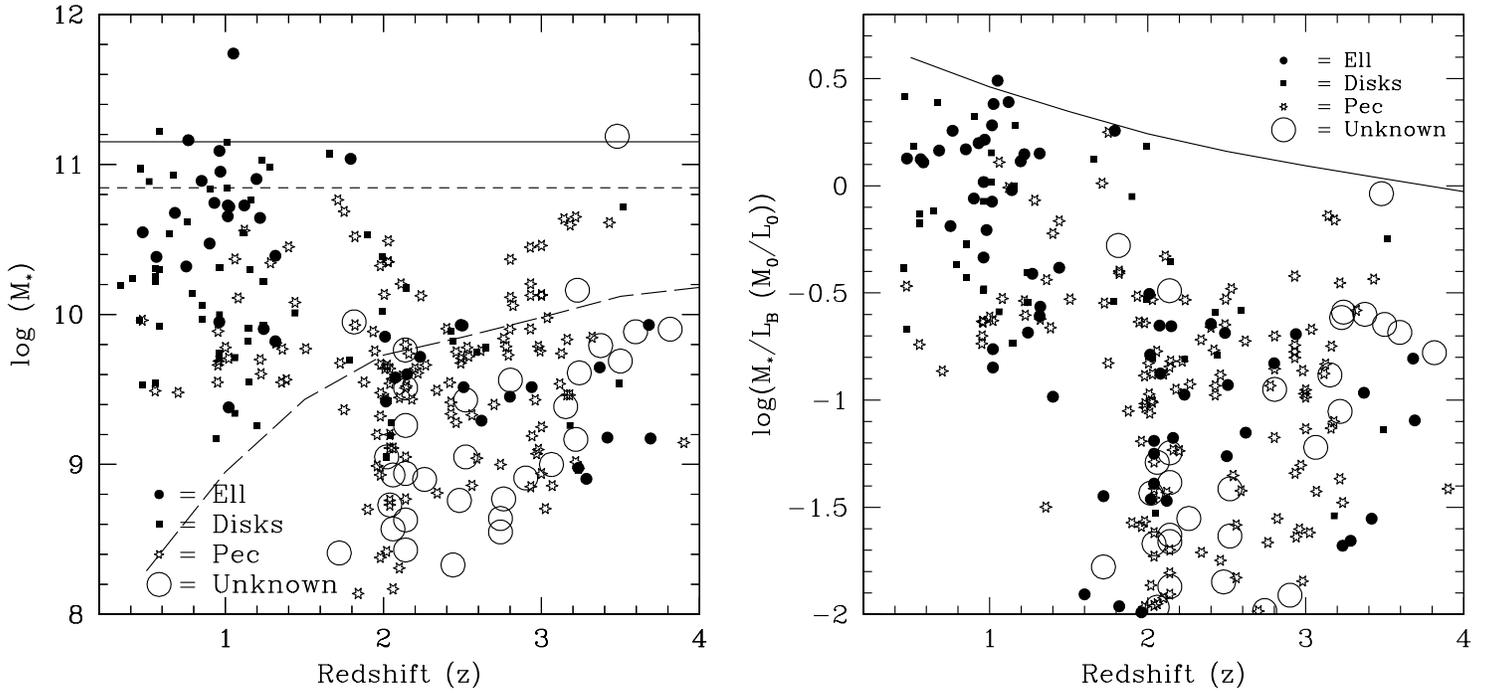}{6.0in}{0}{100}{100}{-310}{-170}
\vspace{0.7in}
\caption{Left panel: The distribution of stellar mass as a function of 
redshift and
morphology.  The solid and dashed lines are the values of
 M$^{*}$ computed by Cole
et al. (2001) for Salpeter and Kennicutt initial mass functions, respectively.
The long dashed line is the curve above which we are complete, corresponding
to a maximally old stellar population.
Right panel: the distribution of stellar mass to light ratios (M$_{*}$/L$_{B}$)
as a function of redshift and morphological type in the HDF fields.  
The solid line
shows the evolution of M$_{*}$/L$_{\rm B}$ for a single stellar population that
is as old as the universe's age at each redshift.}
\end{figure}

\begin{figure}
\plotfiddle{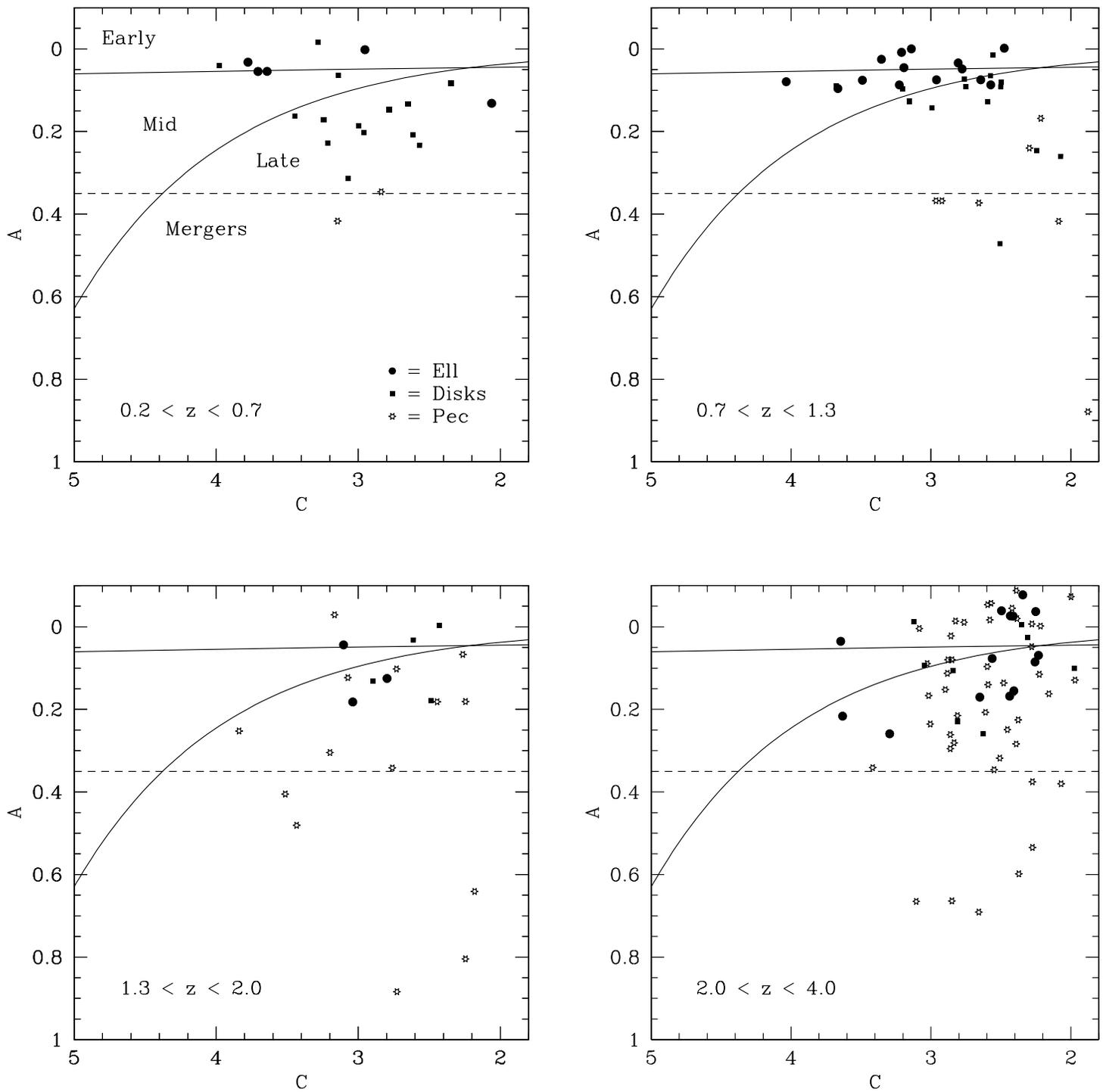}{6.0in}{0}{100}{100}{-310}{-170}
\vspace{0.7in}
\caption{The concentration-asymmetry diagram for galaxies in the HDF-N and
S divided by redshift.  The curved and straight solid lines and the dashed 
line divide spaces where early,
mid, late-type galaxies and mergers are found in the nearby galaxy 
population (Conselice et al. 2000a; Bershady
et al. 2000).  Plotted on this diagram are galaxies which are chosen by
eye to be  ellipticals, disks or peculiars.}
\end{figure}

\begin{figure}
\plotfiddle{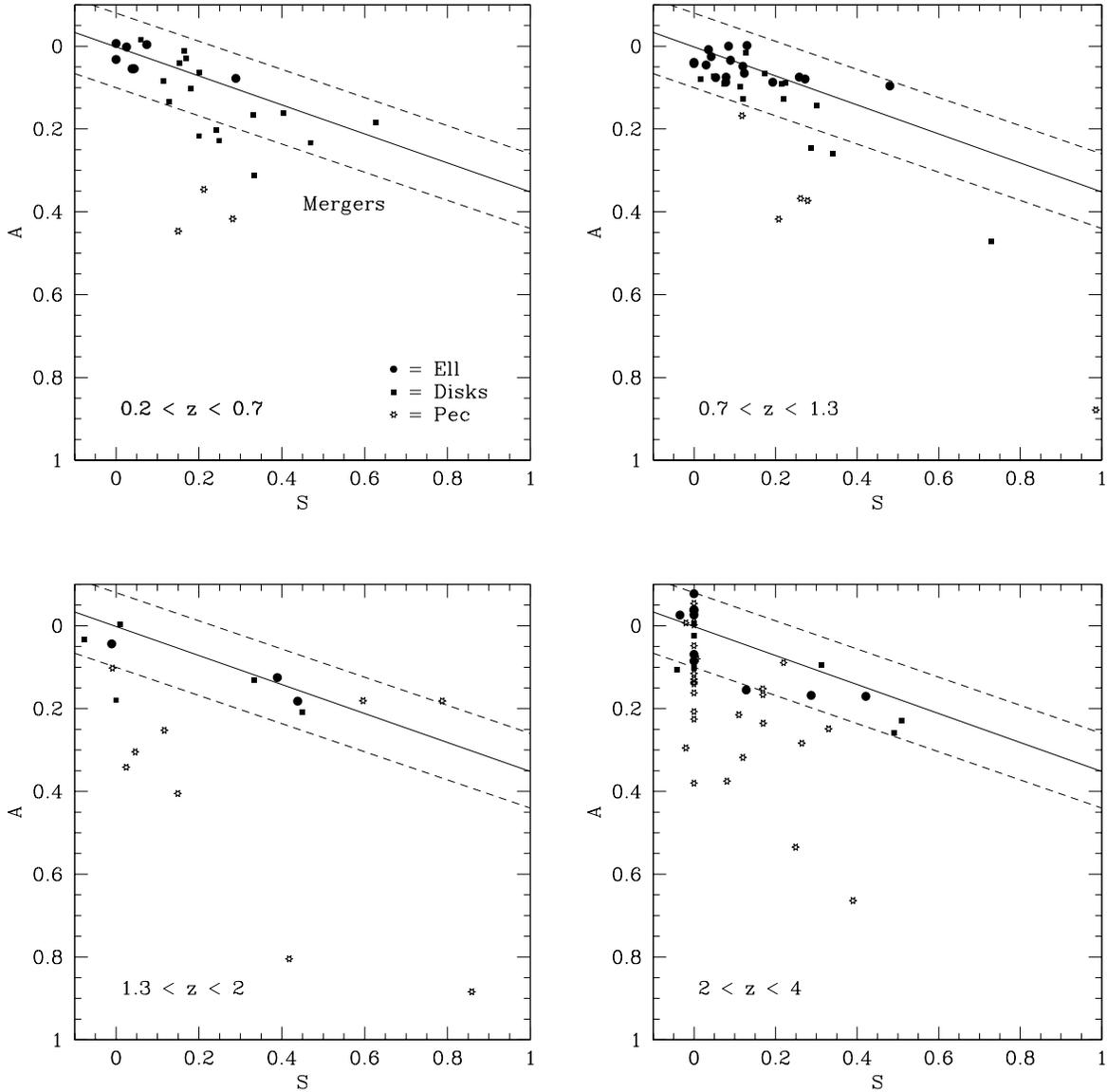}{6.0in}{0}{80}{80}{-270}{-150}
\vspace{0.7in}
\caption{The clumpiness-asymmetry diagram for galaxies in the HDF-N and HDF-S.
The solid line is the relationship between $S$ and $A$ found for
nearby normal galaxies (Conselice 2003), while the dashed lines are
the 3$\sigma$ scatter for this $z \sim 0$ relationship.  Galaxies which
are too asymmetric for their clumpiness are mergers in the nearby universe.}
 \end{figure}

\begin{figure}
\plotfiddle{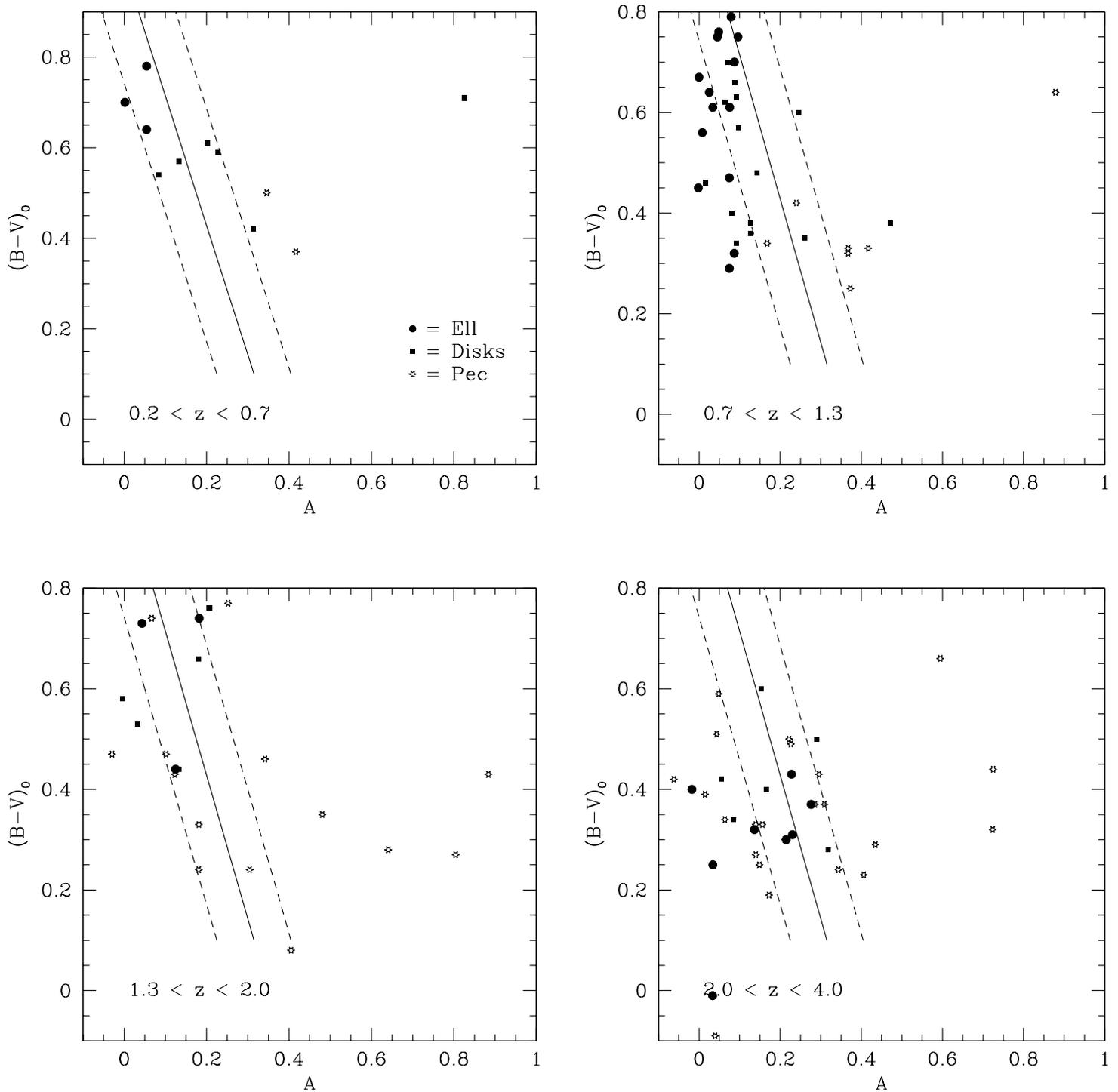}{6.0in}{0}{100}{100}{-310}{-170}
\vspace{0.7in}
\caption{The color-asymmetry diagrams for the HDF fields.  The solid line
is the relationship between these two quantities at $z \sim 0$ (Conselice
2003).  The dashed lines are the $z \sim 0$ 3$\sigma$ scatter from this 
relationship.}
\end{figure}

\end{document}